\documentclass[letterpaper,twocolumn,10pt]{article}
\usepackage{usenix2019_v3}

\usepackage{tikz}
\usepackage{amsmath}
\usepackage{multirow}
\usepackage{array}
\usepackage{tabularx}

\begin{document}

\date{}

\title{Recent Advancements In Distributed System Communications}

\author{
{\rm Ioannis Argyroulis}\\
Vrije Universiteit Amsterdam \\
johnargyroulis@gmail.com
}

\maketitle

\begin{abstract}

Overheads in Operating System kernel network stacks and sockets 
have been hindering OSes from managing networking operations efficiently for
years. Moreover, when building Remote Procedure Calls over TCP, 
certain TCP features do not match the needs of RPCs, imposing additional overheads.
These issues degrade the performance of distributed
systems, which rely on fast communications between machines to be able to
serve a large number of client requests with low latency and high throughput.
The purpose of this literature survey is to look into recent proposals in 
research literature that aim to overcome these issues.
The survey investigates research literature published between 2010-2020, 
in order to include important advancements during the most recent decade at
the time of writing. 
The proposals found in papers have been
categorized into \textit{hardware-based} and \textit{software-based} approaches.
The former require specialized hardware to offer high 
communications performance. 
The latter are implemented in software and don't rely on specialized
hardware or require only certain hardware features.
Furthermore, the proposals where also classified according to whether they implement
\textit{kernel bypass}, to avoid using the Operating System kernel network stack, 
or not.
The hardware-based approaches examined here are RDMA, programmable Network
Interface Controllers (NIC) and
System-on-a-Chip (SoC), while the software-based approaches include 
optimized socket implementations and RPC frameworks, as well as 
user space networking.

\end{abstract}

\section{Introduction}

Communications speed is an important factor in achieving high performance in 
distributed systems, but it is impeded by overheads imposed by
inefficient socket, network stack and RPC implementations. 
Coordination and serving requests require frequent data exchanges between 
machines in a distributed system, making minimizing the cost of these exchanges
essential in achieving less latency and higher throughput in request processing.
Unfortunately, the sockets and kernel network stacks 
provided by Operating Systems (OS) have inefficient implementations that
introduce delays during data exchanges.
Furthermore, when Remote Procedure Calls (RPC), 
a very common paradigm of distributed system communications, 
are built over TCP,
certain TCP features like packet ordering and acknowledgements
can degrade RPC performance too \cite{r2p2-rpc}.
All of these issues 
prevent distributed systems from fully utilizing the
available networking hardware, by increasing the latency and decreasing
the throughput of data transfers.

The scope of this survey is to look into solutions to 
overheads in distributed system communications 
caused by sockets, kernel network stacks and the use
of TCP to build RPC protocols.
Solutions to these overheads are
very important for distributed systems
that perform frequent data exchanges and rely on fast communications
to perform their operations faster.
Examples of such systems are High Performance, Big Data,
distributed transactions or distributed key-value store systems. 
For those systems low latency and 
high throughput in networking operations are essential.
Moreover, the large number of devices that have internet connectivity these
days, along with a large number of applications that use the internet to perform
their tasks, 
has increased the workload for servers greatly.
This has made the deployment
of distributed systems in large scale environments, like datacenters, a very
common practise in order to handle big workloads.
Therefore the survey examines solutions to the issues with sockets, kernel network 
stacks and the use of TCP for RPCs, that can be deployed in such environments. 
It is assumed here that the networking infrastructure is fast and hence the bottlenecks 
are at the end host. Thus networking middleware and any overheads associated with it, 
like overheads from switches, VPNs, NAT-traversals or IP domains are excluded from the 
scope.
Furthermore, application level overheads like serialization and deserialization are
not part of the scope as well.

Given the scope of the survey the main research question is :

\textit{\textbf{"What has been proposed in research literature between 2010-2020 to
overcome communication overheads imposed by  sockets, kernel network stacks and TCP on
RPCs?"}}.

The timeline of the investigated literature spans the decade of 2010-2020, in order
to look into recent advancements in distributed system communications, but also 
gather enough material to gain a better insight on the technologies presented here.

The solutions to the aforementioned overheads proposed in the papers 
were classified into two categories, the \textit{hardware-based approaches} and the 
\textit{software-based approaches}.
Here \textit{hardware-based approaches} refers to
solutions that require specialized
hardware to provide high communications performance.
On the other hand, \textit{software-based approaches} refers to technologies
and techniques that offer software solutions and don't require
specialized hardware built for them or may only require support for
certain hardware features, like virtualization.
Inevitably, the dependence or not on hardware equipment raises a second
research question : 

\textit{\textbf{"Which one is better for distributed systems communications,
hardware or software-based approaches?"}}. 

Moreover, some solutions avoid using the kernel network stack for communications
and thus implement \textit{kernel bypass}, while others do not do this.
Therefore, papers have been further classified 
into \textit{kernel bypass} or \textit{not kernel bypass} solutions.
All hardware-based approaches presented here implement kernel bypass.
The paper categorization is illustrated in Table \ref{categories}.

\begin{table}[ht]
\centering
\caption{Paper Categorization}
\def\arraystretch{1.5}%
\begin{tabularx}{\linewidth}{|c|c|p{2.3cm}|}
\hline
\textbf{Category} & \textbf{Subcategory} & \textbf{Survey Section} \\ \hline
\multirow{2}{*}{Hardware-based} 
    & Kernel Bypass & RDMA, programmable NICs, System-on-a-Chip \\ \cline{2-3}
    & Not kernel bypass & --- \\
\hline
\multirow{2}{*}{Software-based} 
    & Kernel Bypass & User space networking \\ \cline{2-3}
    & Not kernel bypass & Optimized Socket Implementations, 
    Optimized RPC Frameworks \\
\hline
\end{tabularx}
\label{categories}
\end{table}

The rest of this text is structured in the following manner.
First, the methodology of finding and selecting papers will be 
explained. Then a more detailed description of the communication
overheads will follow.
Afterwards, the first research question will be answered by presenting 
first the hardware-based approaches and then the software-based ones, 
mentioning which ones bypass the OS kernel and which do not.
The hardware-based approaches examined here are 
Remote Direct Memory Access (RDMA), programmable NICs
and \textit{System-on-a-Chip} (SoC),
while the software-based ones are optimized socket implementations and 
RPC frameworks, as well as user space networking techniques.
Then a comparison of these two categories of solutions will answer the
second research question in a \textit{discussions} section, 
towards the end of the text.

\section{Methodology Of Survey
\label{method}}

This survey was conducted by reading research papers. These papers were provided
by three means, the supervisor Animesh Trivedi, from the references of other 
papers and Google Scholar. 
The rest of this section provides more details about the sources of the papers
and the criteria of selection.

\subsection{Papers By Supervisor}

At the beginning of the survey, but also at certain times during it,
the supervisor Animesh Trivedi provided papers related to the topic. These
papers were added to the body of the survey and were : 

\begin{table}[ht]
\centering
\caption{Papers By Supervisor}
\def\arraystretch{1.5}%
\begin{tabular}{|c|c|}
\hline
\textbf{Short Title}     & \textbf{Reference} \\ \hline
ScaleRPC & \cite{scale-rpc} \\ \hline
eRPC & \cite{e-rpc} \\ \hline
RAMCloud & \cite{ram-cloud} \\ \hline
DaRPC & \cite{da-rpc} \\ \hline
R2P2 & \cite{r2p2-rpc} \\ \hline
RPCValet & \cite{rpcvalet} \\ \hline
ZygOS & \cite{zygos} \\ \hline
Snap & \cite{snap} \\ \hline
Netmap & \cite{netmap} \\ \hline
mTCP & \cite{mtcp} \\ \hline
Affinity-Accept & \cite{affinity-accept} \\ \hline
TAS & \cite{tas-tcp} \\
\hline
\end{tabular}
\end{table}

\subsection{References Of Other Papers}

References on some papers led to discovering other papers relevant to 
the content. If those new papers fitted the selection criteria described
below, they were added on the survey too.

\subsection{Google Scholar}

Google Scholar was also used for discovering papers related to the survey topic, 
since it searches through various sources of research publications. 
The search results were filtered by the publishing date range 2010-2020.

\subsection{Criteria Of Selection}

Besides being relevant to the research questions, 
these were the other criteria of selecting or rejecting papers:

\vspace{8pt}

\underline{Publishing date between 2010 and 2020} : 
the reason for selecting this date range
was that the survey focuses on recent advancements. Looking only to the latest
publications though (2020) would exclude papers from previous years that are 
still relevant today. Therefore, a decade ending to the year of writing was 
considered as a suitable timeline for staying within the recent developments, 
without going too far in the past. This also allowed concentrating more material
on certain technologies presented here and provided further understanding on
how these technologies developed over time.

\underline{Focus on technology for servers} :
Only technologies and techniques applicable to servers and usable by 
distributed systems were added on the survey. Papers referring to similar 
solutions for switches and routers or other networking equipment, 
even the software counterparts (like software routers), were not included,
as was explained in the introduction for the scope of the survey. 
The reason was to limit the scope of an otherwise very broad topic.

\section{Inefficiencies In Distributed System Communications
\label{ineff}}
\vspace{8pt}

For many years, distributed system communications have been conducted over
socket APIs provided by OSes, while Remote Procedure Calls (RPC), a
very common communications paradigm for these systems, have been built
over TCP for reliable packet deliveries. 
Both sockets and TCP though have been incurring delays that degrade the 
performance of message exchanges between machines. 
Overheads from the use of socket APIs can be attributed to inefficiencies in 
both the socket implementation, but also in the underlying network stack 
that processes packets in the kernel during socket communications. 
The TCP overheads examined here though are not attributed to the protocol itself,
but on the mismatch of certain features it provides, and therefore enforces,
with the semantics and needs of RPCs.
Understanding why these inefficiencies exist is necessary in order to understand 
what the hardware and software-based approaches presented in this text are 
doing to overcome them.

\subsection{Socket Overheads
\label{ineff_socks}}

Sockets are used to establish connections with remote machines and 
to exchange data with them over the network. 
As explained in \cite{megapipe}, \cite{affinity-accept} 
and \cite{mtcp} though, the socket implementations provided by 
Linux lack efficiency on multi-threaded management of 
connections, while they also suffer overheads by the file 
system and the use of system calls to the kernel. 
These issues incur time costs that increase latency and reduce
throughput during data transfers.

Looking into the multi-threaded management of socket connections, there are
two factors that degrade performance. The first is the contention of multiple
threads on a single \textit{accept} queue in a socket, that contains incoming remote
connections and is protected with an exclusive lock, allowing only
one thread to accept a connection at a time. 
This causes contention between kernel threads that enqueue incoming connections
and application threads that remove them from the queue to manage them, 
which also leads to idle times for threads that wait to acquire the lock.
Recent releases of the Linux kernel deal with this though, by allowing multiple
sockets running on different threads to listen on the same socket address with the
\texttt{SO\_REUSEPORT} option,
while the kernel takes care of load balancing connections on the different threads.
The second factor is the lack
of affinity to the same core for all activities related to a given connection, 
like between packet delivery and reading or writing data on the connection's socket.
There are
technologies like Receive Side Scaling that can enable Network Card Interfaces
(NICs) to load balance incoming connections to threads running on different CPU
cores and keep sending incoming packets of a connection to the same core. 
That affinity though can be lost when an application thread residing on a different 
core accepts a connection or writes data to the socket.
This lack of affinity causes inefficient cache usage
when different CPU cores need to maintain the state of the same connection
and perform different activities on it.

Performance issues can also occur by the association of sockets with the 
file system.
For instance, allocating a new file descriptor for a socket can become 
expensive with a large number of connections. Specifically,
Linux searches for the minimum available file descriptor number 
for the process, which incurs an undesired delay when the application
manages many sockets. Additionally, this searching process uses a lock,
since files are shared within the process, which can cause contention
on a multithreaded environment. Another issue is that sockets are treated
like file types by the Virtual File System, which associates them with
data that is globally visible and subject to system-wide synchronizations.

\begin{figure}[h!]
  \centering
  \includegraphics[width=210pt]{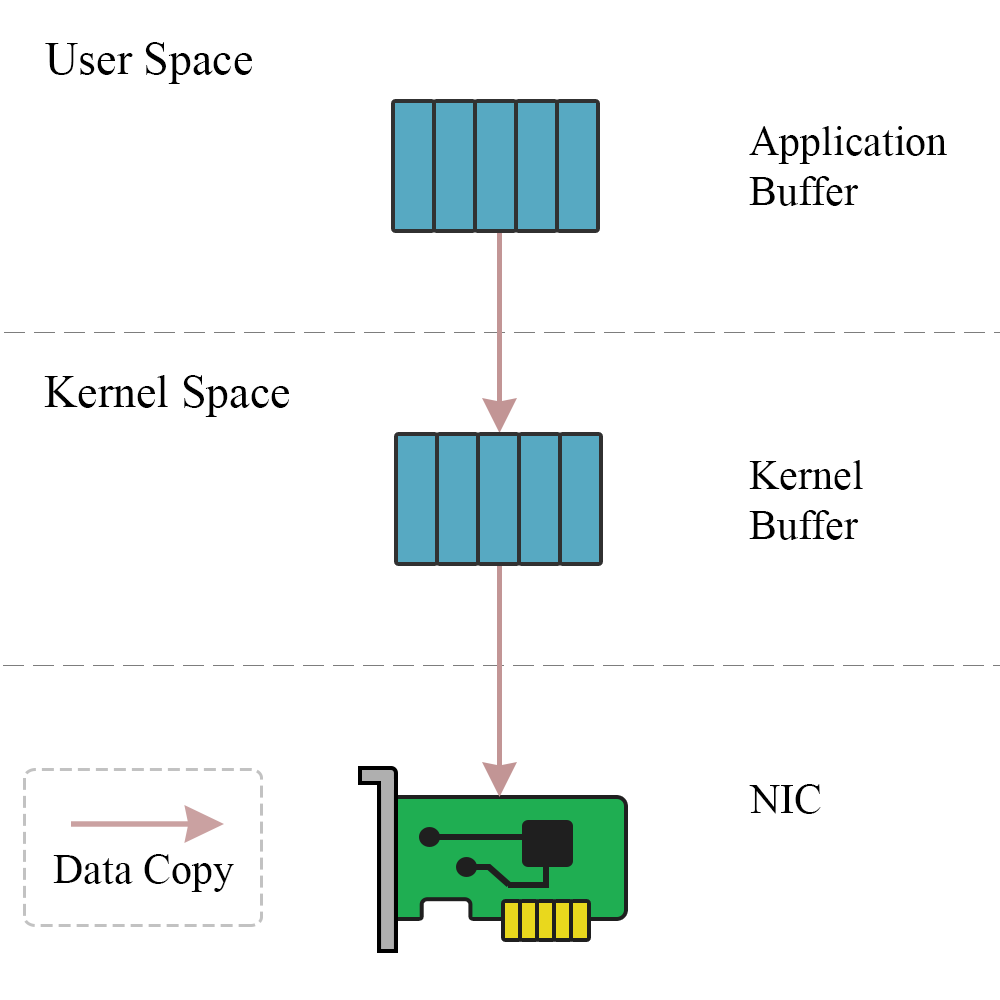}
  \caption{Message data is copied between user space and kernel space 
  buffers when sending a message with sockets. The opposite flow
  happens for message reception.}
  \label{fig:kernel_copies}
\end{figure}

Sockets also use system calls to the OS kernel that incur expensive 
context switches between user space and kernel space and also require
data copies between those spaces.
When an application uses sockets, system calls transfer the control
from the user space application to the kernel, which interacts with the NIC
in order to perform the networking operation requested by the application.
This context switch from user space to kernel space introduces latency. 
Moreover, any data to be transmitted through
a socket is copied from an application owned buffer to another buffer 
inside the kernel, associated with the socket. From there, the data is
finally forwarded to the NIC for transmission over the network. This is 
illustrated in Figure \ref{fig:kernel_copies}.
Reception follows the opposite path, from the NIC to the kernel and then
back to the application, incurring a context switch and a data copy 
between kernel and user space, as with transmission.
These context switches and data copies delay socket operations and 
decrease the throughput in request processing in distributed systems.

\subsection{Kernel Network Stack Overheads
\label{ineff_kern}}

The OS kernel's network stack's processes incoming and outgoing
packets and implements networking protocols like TCP and UDP.
The network stack implementations on OS kernels though have been
unable to process incoming packets at the full speed of network links for
years. The problem lies in the time it takes to process a packet
in the stack. When receiving packets from a high speed network link,
like 100 Gbps, the kernel network stack has very limited available time
to process the packets, in order to achieve a processing rate which is 
the same as the link rate and fully utilize the link capabilities. The
same can be observed even with slower links when small packets are sent,
because small packets can be transferred by NICs at high rates as well.
The delays in packet processing put a limit on the throughput of OS
network stacks that also limits the maximum achievable performance in 
distributed system communications.

These delays are attributed to the code of kernel network stacks, which
incurs high computational complexity and memory footprints \cite{tas-tcp}. 
Network stack processing has a modular design with various stages, 
connected with function calls, queues and software interrupts, built for
extensibility. This design though results in a large number of
instructions per packet, while software interrupts exacerbate the processing
delays.
The TCP state machine is also implemented in a monolithic code block with 
many branches that does not leverage batching and prefetching efficiently, 
causing cache mispredictions and increasing the instruction cache memory
footprint.
Moreover, inefficient cache usage results from the use of large metadata
structures that span multiple cache lines and can cause false sharing, 
while state might be shared by multiple cores requiring locks and causing
cache coherence overheads. 
Additionally, expensive dynamic memory allocations for packet buffers and
metadata per packet happen frequently during data transfers. 
All these issues with network stacks aggregate time costs that
make it hard to achieve packet processing at link rates.

\subsection{TCP Overheads On Remote Procedure Calls
\label{ineff_tcp}}

Building Remote Procedure Calls on top of TCP 
for reliable packet delivery can actually
hurt the communications performance of RPCs \cite{r2p2-rpc}. 
One of the causes for TCP's mismatch with the requirements of RPCs, is
its design to perform better with large data transfers
spanning multiple packets, 
while RPC messages are often small and can fit to a single packet.
For such small RPC packets, 
that are considered either requests or responses and
are semantically independent,
TCP's packet ordering is unnecessary and only 
adds extra delay.
Furthermore, RPC responses are also acknowledgements
themselves, so sending separate packet acknowledgements with TCP also adds
unnecessary delays.
Moreover, by establishing only one-to-one connections, TCP requires 
maintaining and managing a lot of connection state, which becomes harder
as the number of connections increases, causing a scalability limitation.
To avoid such a limitation multiple RPCs are often layered 
on top of the same TCP connection. This
can also lead to head-of-line blocking between different RPCs,
while scheduling options are limited.

After explaining the performance issues with traditional communication
methods for distributed systems and their causes, the following sections 
continue with
hardware and software-based approaches to overcoming them.
Among those approaches there is a subset that 
completely bypass the kernel by either offloading networking
operations to specialized hardware or creating their own
network stack implementations as software run in the user space.
The term \textit{zero-copy transfers} is also mentioned a lot with 
\textit{kernel bypass} techniques and here it means that there are no
intermediate
copies of transmitted or received data between user space and kernel space 
buffers, unlike with OS-provided sockets.
Instead, applications and NICs share the message buffers and the time cost
of intermediate copies in the kernel is eliminated.
The RDMA technology presented in the next section is a hardware-based
approach that can achieve very low latency and high throughput by 
implementing kernel bypass and zero copy transfers.

\section{Remote Direct Memory Access}
\subsection{Overview}

Remote Direct Memory Access (RDMA) is a feature that allows an 
application to read or write memory located on a remote machine, 
without requiring the active participation of an application in that 
remote machine during the data transfer. 
The RDMA technology though supports both the RDMA feature and also
messaging semantics, in which a receiver in the remote side is notified
about a new message and thus participates in the communication.
In the rest of this text RDMA refers to the technology and not the
feature.
With the RDMA technology
applications can specify memory regions that will be used during 
the communications, before the latter commence.
Then RDMA-capable Network Card Interfaces (NICs) can use Direct Memory Access (DMA)
to directly read or write data on these memory regions without involving
the kernel.
Additionally, applications can post transmission or reception requests 
directly to the local NIC, without passing through the kernel.
After the NIC receives a request, it can use DMA to either fetch local data 
that will be transmitted or to store remote data locally, when it receives it 
from a remote NIC.
DMA does not involve the kernel and thus does not incur context
switches and additional data copies between kernel and user space. 
In this manner, the kernel is bypassed and zero-copy transfers are ensured,
as the NIC is the one performing the networking operations, thus saving
CPU cycles for other tasks.

With RDMA an application can choose two ways to communicate with another 
machine. 
First, as already mentioned, it can either read or write
directly on a memory buffer in the remote machine without notifying the 
remote application.
This is useful in cases where there is no need for the remote side to 
perform some task upon the reception of a message. 
Second, RDMA also supports sending messages for which the receiver 
is notified on arrival, which can be used when the receiver has to 
act upon the reception of a message, like in RPCs.
Both options are available through RDMA's communication primitives, 
which send requests directly from the user application to the NIC.
Moreover, RDMA provides connection methods that can offload operations 
like reliable transfer and packet ordering on the networking hardware,
allowing the CPU to perform other tasks.
With kernel bypass, zero-copies and offloading certain networking 
operations to hardware, RDMA can achieve lower latency and higher 
throughput than traditional socket based networking.

Technologies that support RDMA are Infiniband, RoCE and iWARP. 
Infiniband uses specialized networking equipment that
provides a high speed lossless network (packet errors are rare) and
implements flow control mechanisms.
RoCE and iWARP on the other hand, support RDMA semantics over Ethernet 
with the use of hardware adapters. 
One of the differences between RoCE and iWARP is that the latter is
designed to provide RDMA over TCP/IP, by offloading TCP operations 
to the NIC. RoCE version 2 on the other hand uses UDP/IP and requires
a lossless network to provide low latency.
By offloading certain operations to the 
networking hardware, like reliable delivery, flow control and ordering, 
these three RDMA technologies save CPU cycles for other tasks. 
There are also ways to use RDMA entirely in software like Soft-RoCE and
SoftiWARP,
but they do not offload operations to hardware like the previous 
options and thus achieve less CPU efficiency and lower performance.
Therefore, in production environments like datacenters, Infiniband,
RoCE or iWARP are more suitable to achieve the best RDMA performance 
and this is the reason for categorizing RDMA to the hardware-based 
approaches of the examined literature.

RDMA has gained a lot of popularity over the years for its low 
latency of a few \textit{$\mu$sec} and high throughput of millions of
operations per second.
RDMA's high performance has already led to its adoption in a variety of
distributed systems, like key-value stores, distributed consensus and
transactions, High Performance Computing and Deep Learning systems.
Some systems have been designed from scratch to work with RDMA, 
while others have been extended to support it, 
like in \cite{hadoop-rpc} for the RPC communications in the Hadoop 
ecosystem, in \cite{memcached-rdma-kv} 
for boosting the performance of the Memcached key-value store or in
\cite{ar-grpc} for optimizing gRPC for tensor transfers in Tensorflow.
To gain a better understanding of how RDMA is used in distributed
systems, it is important to explain in more details the networking
operations and connection methods that it supports.

\subsection{Networking Operations 
\label{rmda_net}}

\begin{figure}[h!]
  \centering
  \includegraphics[width=210pt]{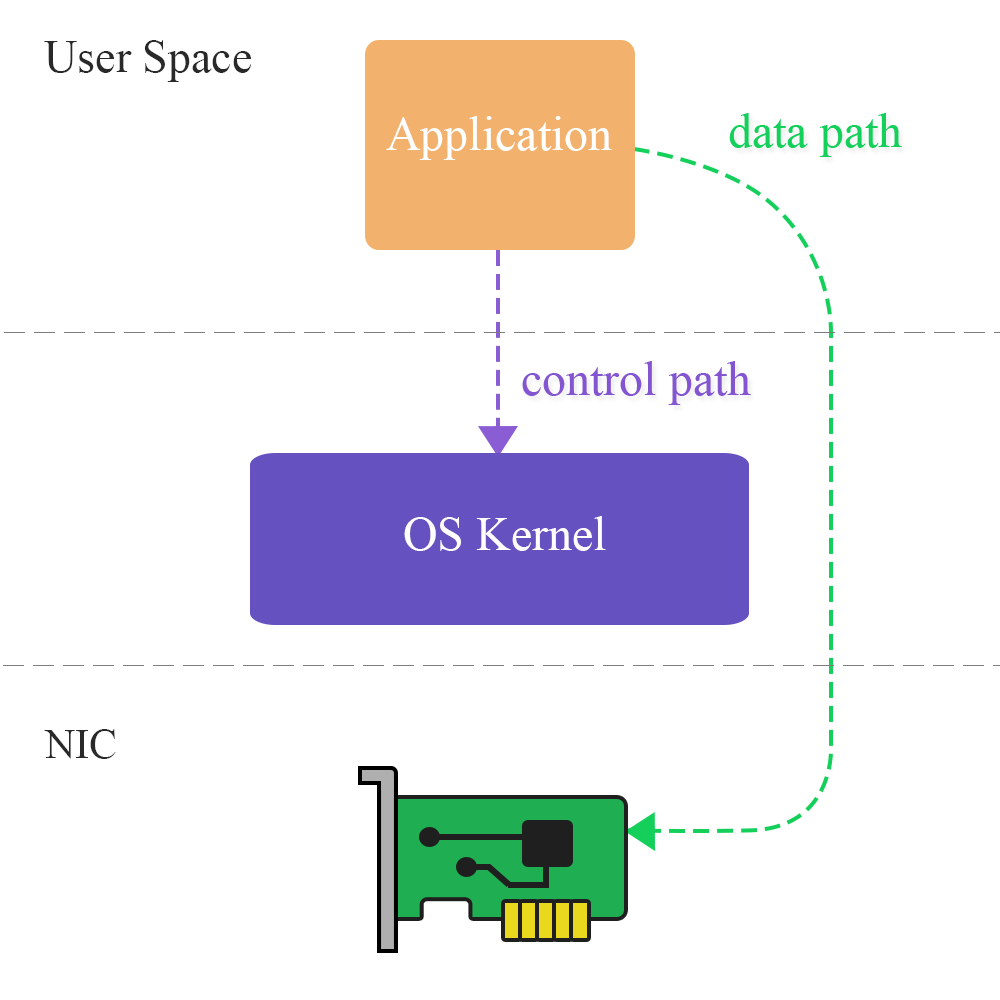}
  \caption{Control and data paths in RDMA. The control path uses
  the OS kernel, while the data path bypasses it.}
  \label{fig:rdma_control_data}
\end{figure}

In RDMA the operations that take place to accomplish data transfers 
are split into two types, those belonging to the \textit{control path} 
and the rest belonging to the \textit{data path}. 
The control path contains operations related to managing the resources 
for the communications, like pre-allocating message buffers and registering
them to NIC to be used during data exchanges. 
Such operations are slow, but happen before commencing the communications 
and hence their overheads do not have to be repeated with each data 
transfer, as with sockets. 
This requires knowing beforehand how many resources to allocate, while
with sockets memory allocations happen dynamically.
On the other hand, the data path operations take care of transferring 
the data to its destination and copying it from a network card to 
memory or the opposite. 
The control path uses the kernel to manage resources while the data path 
is the one that uses Direct Memory Access to bypass the kernel and reduce 
CPU usage, as illustrated in Figure \ref{fig:rdma_control_data}. 
This separation of paths that avoids repeating the time cost of the 
control path in every data transfer reduces latency and leverages the 
saved time to increase the rate of networking operations performed.

RDMA provides its own primitives to communicate with a remote machine, 
which are also called "verbs". 
There are two types of such network operations, the 
\textit{two-sided} and the \textit{one-sided}, which have also been
called \textit{messaging} and \textit{memory} verbs.
The \textit{messaging verbs} are the SEND and RECV operations that
send and receive data respectively. The are two-sided operations
because both sender and receiver participate in the communication.
The receiver has to post a RECV operation to the NIC prior to receiving 
messages and it is notified upon reception by the NIC.
The \textit{memory verbs}, or one-sided operations, perform data transfers
that do not involve the remote side. There is a WRITE operation that
can directly write data to a memory location on a remote machine and
a READ operation that directly fetches data from a remote memory 
location. With one-sided verbs the NIC doesn't notify any application 
thread on the remote side that a data transfer took place.
One-sided verbs also include the remote atomic operations 
\textit{fetch-and-add} and \textit{compare-and-swap}, which can
guarantee atomic access on remote memory locations, as long as this 
access happens between NICs on different machines.

Both SEND and WRITE can optionally transmit a small \textit{immediate} value 
along with the data, which is sent to the remote side as a notification.
If this option is used, then WRITE will involve the remote side to process this
value. Although it seems like a minor detail, the 
\textit{immediate} value has been used in constructing 
request-response or RPC protocols with one-sided operations.

\subsection{Connection Methods}

RDMA also supports different types of connections between machines, 
which play an important role in the design of a distributed system.
Before describing these connection types and their differences though, it's necessary to introduce \textit{Queue Pairs} (QPs). 
RDMA communications are established through Queue Pairs, which contain
a Send and a Receive Queue. 
To perform an RDMA operation, an application has to submit a Work Request (WR) to a QP; the WR describes the RDMA operation to perform and which
network buffer to use. Additionally, 
a submitted WR is stored as a Work Queue Element (WQE) 
in either the Send or Receive queue of the QP, 
depending on the type of RDMA operation. The RDMA NIC executes the 
operations described in the WQEs of QPs.

Having explained QPs, the different RDMA connection types can
be introduced. These are Reliable Connection (RC), 
Unreliable Connection (UC) and
Unreliable Datagram (UD).
With Reliable Connection a QP is associated with only one other QP.
Messages are delivered by the NIC reliably by using acknowledgements 
and packets are ordered. RC is similar to a TCP connection, but reliability
and packet ordering are managed by the NIC.
With Unreliable Connection a QP is again 
associated with only one other QP, but the connection is \textit{unreliable},
which means packets are not ordered, may be lost and no acknowledgements
are sent. 
Applications have to take care of retransmissions themselves when 
errors occur during transmissions.
Finally, with Unreliable Datagram a QP can send and
receive single packet messages to/from any other QP and multicast is 
also supported, which is not available in other connection types. 
UD is unreliable like UC and similar to UDP 
sockets.

Different types of connections support different networking operations and 
maximum message sizes. The following table is taken from Mellanox's
user manual \cite{mellanox-guide} to show what can each type of 
connection support. The MTU entry in the last row stands for 
Maximum Transmission Unit.

\begin{table}[ht]
\centering
\caption{RDMA Connection Methods - Feature Support}
\def\arraystretch{1.5}%
\begin{tabular}{|c|c|c|c|}
\hline
\textbf{Operation}     & \textbf{UD} & \textbf{UC} & \textbf{RC} \\ \hline
Send (with immediate)  & X           & X           & X         \\ \hline
Receive                & X           & X           & X         \\ \hline
Write (with immediate) &             & X           & X         \\ \hline
Read                   &             &             & X         \\ \hline
Atomic Operations      &             &             & X         \\ \hline
Max. Message Size      & MTU         & 1GB         & 1GB       \\ \hline
\end{tabular}
\end{table}

\section{Using RDMA Efficiently}

Using RDMA to achieve high performance communications is not as
straightforward as switching to a new API. There are important factors
that affect the performance, such as the choice of the RDMA
connection method, taking actions to improve scalability,
deciding whether to use one-sided operations or
Remote Procedure Calls (RPC) and deciding whether to use an existing RPC
framework or implement a custom RPC protocol. 
Therefore, distributed system designers need to understand
these factors in order to make choices that match their system's
requirements and performance goals. To provide this understanding,
the rest of this section discusses how these factors impact the
performance of distributed system communications with RDMA.

\subsection{Choosing The Right Connection Method 
\label{rdma_connections}}

Choosing the right connection method to fit a system's needs requires
comparing their features. As already shown,
the three methods support different subsets of RDMA networking operations 
and offload different functionality to the NIC. Thereby, some of them 
can satisfy a system's requirements better than others. 
Moreover, all three connection methods have benefits and pitfalls that 
are not obvious by their features alone and users 
of RDMA have to be aware of them before choosing a connection type.

A Reliable Connection (RC) is very useful when reliability and packet
ordering guarantees are necessary, like with a TCP connection. 
By offloading these two operations to the NIC, RC allows the CPU to 
save cycles for other tasks, leading to better overlapping of 
computations with communications. This overlapping can be improved when
running an RC connection over an Infiniband lossless network, which takes
care of flow and congestion control with the networking equipment as well.
Moreover, RC supports all networking
operations, giving the system designers more freedom of choice in how to
perform data transfers or message exchanges.
A drawback of RC though is that, like
TCP, it supports only one-to-one connections and managing this connection
state with a large number of clients can lead to a performance 
degradation. This scalability issue though can be solved and techniques to
do so will be presented in the following subsection. It would suffice to 
say for now that ensuring scalability requires taking certain steps
during building the RDMA communications layer of a distributed system, 
which increases the complexity of the system design process.

An Unreliable Connection (UC) doesn't provide reliability and ordering
guarantees, reducing the processing time on the NIC and the network 
traffic by not sending packet acknowledgements. 
Hence it allows the NIC to spend more time in handling requests 
than dealing with these guarantees when they are not required.
As explained in R2P2 \cite{r2p2-rpc}, packet ordering is often redundant for 
Remote Procedure Calls, as requests and responses usually fit in a single 
packet and they are semantically independent.
Furthermore, UC is also useful when 
sending acknowledgements for received packets is not necessary.
More specifically, lossless networks lose packets rarely, 
so running UC over such a network will deliver the packets reliably most 
of the time, without the need for the NIC to send acknowledgements.
Moreover, in an RPC protocol, the response can also act as an acknowledgement, 
thus the NIC does not need to send one.
A drawback of UC though, is that although rare, any occurring errors and
retransmissions have to be handled with software, which adds extra
complexity on the software development process and involves the CPU when it
happens.
Additionally, UC supports only the two-sided 
operations and WRITE, hence systems cannot use UC to read remote data
directly with READ or use the atomic RDMA operations. Finally, since 
UC supports one-to-one connections like RC, it suffers from the same
scalability issue.
    
Unreliable Datagram (UD) bares similarities with UC but comes with its 
own advantages and disadvantages as well.
Like UC, it doesn't offer packet reliability and ordering and 
can be useful in cases where these guarantees are not necessary, as
explained before. Additionally, its support for multicast makes it suitable 
for systems that require this capability.
Moreover,
UD can offer greater scalability than RC and UC, since a UD Queue
Pair can be used to exchange messages with any other UD QP. There is no need for a server to have a separate
QP for each different destination. Therefore,
less connection state has to be maintained
than using RC and UC connections. Furthermore, UD also makes it possible 
to batch requests for networking operations to the RDMA NIC
regardless of the destination of the messages, 
while RC and UC support only batching such requests 
for the same destination \cite{fasst-tr}.
One limitation with UD though is that it
supports only two-sided operations, which always include the step of
notifying the remote side. Another drawback is that like UC, it requires 
developing code to handle retransmissions when errors occur.

Choosing the best suited connection method is not restricted to picking one
of the three, as combinations of them are also possible.
For example,
the HERD distributed key-value store \cite{herd-kv} uses both UC and 
UD in their RPC protocol, in which multiple clients communicate with 
one server.
The clients send their requests to the server with UC, while the server 
responds through a UD connection. 
UC allows the clients to write their requests directly to 
the server's memory with one-sided WRITE and the server polls the memory 
location to detect new requests. Although this could also be implemented
with RC, UC is preferred in this case because it involves less processing in 
the NIC, and reduces the network traffic by not sending packet delivery
acknowledgements.
Moreover, the scalability issue with UC is greatly mitigated by the fact
that the server does not use it to reply to clients. Although there are
multiple UC connections from clients to the server, 
the server only uses the UC connections to receive client requests.
Thus, the server's NIC manages less connection-related state than if UC was also
used for sending responses. 
For client responses, the server uses UD to ensure better scalability than UC. 
For systems
that have different needs in different parts, like HERD, such a combination
can lead to more in-depth optimization and hence, higher performance.

As it should be obvious by now, there is no perfect connection type, as
they all have their benefits and drawbacks. Distributed system designers
have to decide on which one to use based on which features are desirable,
but also which tradeoffs are tolerable. Different connection types can 
also be used in the various parts of a system, to further optimize the
data exchanges, according to the needs of each part.

\subsection{Improving Scalability On RDMA Communications 
\label{rdma_scalability}}

RDMA does not guarantee scalability by default
and users have to be aware of what can cause scalability issues and 
what can be done to deal with them. Scalability problems occur 
from thrashing 
in the limited cache on the RDMA NICs that stores connection state, high 
memory consumption by allocating a message buffer for every connection, 
and allocating and managing multiple Completion Queues (CQs). 
Completion Queues store notifications about completed Work Requests
submitted to QPs. Two-sided SEND/RECV uses CQs for notifications 
about new message receptions.
The rest of this section presents solutions proposed for the 
scalability issues with RDMA communications, found in research
literature.

\subsubsection{Dealing With Limited NIC Caches}

As mentioned in \ref{rdma_connections}, 
RC's and UC's performance degrades with a 
large number of connections. 
This happens because RDMA capable NICs contain 
caches with limited amount of memory, that can store state related to 
active connections.
The cached data is actually comprised of Queue Pair (QP) states, 
Work Queue Elements (WQE), Memory Translation
Table (MTT) entries and Memory Protection Table (MPT) entries.
MTTs store virtual to physical memory address translations 
and MPTs store memory access permissions.
When the aforementioned data does not fit into the NIC's cache, 
it has to be fetched from memory. This results to slower data retrieval
and degrades performance with many active connections.

One way to reduce cached state in the NIC is to minimize the cached MTT and 
MPT entries. This can be done by using either Linux huge pages or registering 
physical pages directly to the NIC. The MTT and MPT entries refer to 
memory locations registered to the NIC, used during 
communications to store messages. 
With Linux huge pages, it is possible to register large
contiguous segments of memory that are not swapped. Therefore, to cover the 
memory needs of an application, using huge pages leads to fewer but larger 
segments of memory, instead of using more and smaller ones with the default 
Linux page size. 
As a result, less MTT and MPT entries have to be maintained.
Moreover, smaller objects can be allocated and managed inside the large memory 
regions. FaRM \cite{farm}, a distributed key-value store,
registers 2 GB pages for its RDMA operations because of its large memory needs.
A drawback of using very large pages like FaRM though, is that it can lead a 
lot of fragmentation on the system's memory space.
If this happens, the larger the pages, 
the less possible it will be to find a contiguous 
memory region of the requested size to allocate a page.
One the other hand, some RDMA NICs also support registering physical 
pages directly, which eliminates the need to store virtual to physical 
address translations, reducing the information to be cached in the NIC
as well. 
LITE \cite{lite}, an indirection tier for RDMA in the Linux kernel, 
registers the whole available memory to the NIC with this method, but 
offers higher level memory management and smaller object allocations
on top of the registered memory.

Reducing the QP state that is cached in the NIC is another important step to
achieve greater scalability. 
As previously explained, UD leads to greater scalability than 
point to point connections, since it maintains less QP state.
When using RC or UC instead, which require point to point
connections, another way to reduce QP state
is to share QPs among groups of 
threads like the FaRM \cite{farm} key-value store does. 
Since RC and UC 
establish a connection by associating one QP with only one other QP, 
this approach is actually sharing the connection between threads. 
Therefore, it can aid in avoiding establishing multiple connections 
to the same destination between different threads.
Moreover, FaRM's technique requires synchronization, 
since multiple threads access the same QP. By grouping the threads and
allowing a QP to be accessed by only one group, the synchronization 
overhead can be mitigated, since there is less contention for the same 
QP. On the other hand, ScaleRPC \cite{scale-rpc}, an RPC framework over RDMA
that uses RC, reduces the QP state that resides in the NIC's cache by
grouping client connections instead of threads and serving only one group at 
a time with time slices. Its goal is to be able to serve as many 
connections as possible during a time slice, without incurring cache thrashing 
in the NIC. So instead of reducing the overall number of QPs maintained, 
ScaleRPC focuses on limiting the number of connections 
that are served at a time to minimize cached QP state, but
without sacrificing throughput.

\subsubsection{Minimizing Memory Consumption And CQ Polling Overheads}

Besides the NIC's caching issues, two more problems occur with multiple
connections. One is 
that creating and managing message buffers for every connection leads
to a growth in memory consumption with the number of connections.
The memory limits then also affect the maximum number of connections that
can be served.
The second problem is that a connection is associated with a Completion Queue
(CQ) when two-sided RDMA operations are used, which stores notifications 
about the reception of messages. Maintaining multiple such queues in memory
and polling them for completions can also pose a burden on the system which 
gets worse as connections increase.

Minimizing the message buffers can help in expanding the boundary of how 
many other machines can a server communicate with. 
To allocate less message buffers on the receiver's side, 
RDMA's Shared Receive Queues (SRQs) can be used. 
SRQs allow many QPs to be associated with one receive queue that stores 
the messages that are received. Since QPs represent different connections,
this method is like sharing a memory region allocated to store messages
among connections.
Another option is to share receive message buffers by groups 
of threads, like in GRAM \cite{gram}, a distributed graph processing engine.
In GRAM a thread can send a message to another thread in a remote location.
Connecting every thread on a server with every other thread on a remote server
and doing this for many servers can lead to allocating and managing a 
large amount of message buffers.
Although GRAM supports this method of one-to-one connections between threads,
it can also create groups of threads and share receive message buffers
between them.
This approach requires demultiplexing the received messages to their 
destination threads. 
To also reduce the number of the message buffers for storing messages to 
be sent to a remote server, GRAM assigns to each server thread only a 
single sending buffer for messages towards all the threads of the a remote
location.
This allows batching the messages of a thread towards the other server, but
requires demultiplexing those messages on the remote side to reach the 
correct receiving thread. 

The memory-bound scalability problem with message buffers is the same 
for Completion Queues, that store notifications about
Work Requests that have been completed by the RDMA NIC. 
Applications detect notifications in Completion Queues (CQ) 
either with polling or blocking until new notifications are
available. 
As a reminder, CQs also store notifications about received
messages when two-sided SEND/RECV is used.
If each RDMA connection is assigned a CQ, 
the larger the number of connections,
the more memory is required for CQs. Moreover, 
if notifications are 
detected with polling, then the more the CQs, the harder it
becomes to poll all of them for notifications.
It is possible though to share CQs among connections, which is very 
convenient in saving memory space and reducing overheads when
polling the CQs. 
For instance, DaRPC \cite{da-rpc}, is an RPC framework based on RDMA, that 
uses this method to create 
a pool of CQs and share a CQ with a subset of the connections.

Combining all the aforementioned techniques to overcome scalability issues can aid in maintaining 
salable performance with RDMA communications, but the networking equipment 
has been improved over the years to aid in this effort too.
In Storm \cite{storm-tran} and DrTM+H \cite{drtm+h}, 
the authors claim that newer NICs
support larger caches and optimizations that deliver high throughput, 
even with cache misses.
Moreover, modern NICs contain powerful processing units that can handle
requests in parallel, hiding the latency of data fetching 
on cache misses with parallelism. Hence, it is expected that as the RDMA
hardware keeps evolving over the next years, distributed system designers 
that wish to use RDMA will have to worry less about dealing with cache 
misses on the NIC and more about managing the message buffers and 
completion queues efficiently.

\subsection{Choosing One-sided RDMA Or RPC}

After selecting a connection method and ensuring scalability,
the third factor that affects the communications performance of a
distributed system is choosing between one-sided RDMA operations and
Remote Procedure Calls.
Each of these choices offers benefits that 
the other can't offer and as with connection methods, selecting the 
right one depends on the system's needs.
Moreover, this dilemma exists for both data transfers 
and remote locks over RDMA, and further explanation follows, regarding
each of these two use cases.

\subsubsection{One-sided RDMA Versus RPC For Data Transfers}

There are two main use cases that point to choosing one-sided RDMA
operations instead of RPCs for a data transfer. 
The first is when there is no need to notify 
the remote server of the data transfer. The second is when there is no 
need for the remote server to process the received data and send back a 
response.
For example, the DARE \cite{dare-cons}
distributed consensus system uses the one-sided WRITE to write "heartbeat"
messages to the memory locations of replica servers. 
These replica servers check for heartbeats periodically to detect failures,
so they don't need to be notified about the "heartbeat" at the time that it 
is updated. Moreover, no response is required and thus, 
there is no need to use an RPC for this case. 
On data transfers that do not need an action from the remote side, like
when replicating information, one-sided operations are obviously the
perfect candidate, since they don't involve the remote side at all.
For instance, in DARE, 
the leader uses READs and WRITEs to directly read and update remote 
consensus logs respectively. 
The DrTM+H \cite{drtm+h} distributed transactions system also updates remote 
transactions logs this way.

Additionally, when the remote memory locations of the data transfers 
and the sizes of the data to transfer are 
known in advance, one-sided operations can possibly replace RPCs.
Distributed key-value 
stores like \cite{pilaf-kv}, \cite{memcached-rdma-kv}, \cite{hydradb-kv}, 
\cite{cell-kv} use READ to directly retrieve the requested 
values of keys from a memory location, when the locations of these values
are known. When they are unknown, these key-value stores either traverse the
index data structure with multiple READs or use RPCs.
Afterwards, they cache the actual location of the desired value,
in order to perform subsequent lookups to these memory locations 
directly with READ, avoiding the repetition of the searching cost.
In \cite{dist-learn}, one-sided RDMA operations replace RPCs for 
exchanging tensors in Distributed Deep Learning. 
The reason that RPCs are substituted
by one-sided operations in \cite{dist-learn}, is that RPC frameworks 
may implement intermediate buffering internally, which leads to data
copies between application and RPC framework buffers and increases
latency.
RPC frameworks may use internal buffers due to not knowing the memory
locations that their users wish to store the messages beforehand.
Additionally, the framework buffers may be small for memory consumption
scalability reasons, which can lead to splitting large 
messages into smaller segments to transfer.
In RDMA communications in \cite{dist-learn} though, there is no need 
for using intermediate buffers.
Senders and receivers exchange remote memory locations where tensors
will be saved. Moreover, tensors shapes are either fixed during computations, or they can be dynamic and
change during computations. In the first case the memory to allocate
for tensor transfers is known before starting the computations. 
In the second case, the sender only sends metadata to the receiver, 
from which the receiver learns how much memory to allocate for the
tensor data and which memory location on the sender has the data, 
and then uses READ to fetch the tensor data from the sender.
Therefore, intermediate buffer copies of an RPC framework are avoided
in \cite{dist-learn}.

On the other hand, there are cases where the completion of the data transfer
requires an action from the remote side and a response is needed by the sending 
side, which makes RPC the best candidate to implement this functionality.
RPCs can allow the remote server to perform operations that are not possible or
easy to do with only READ/WRITE, like moving, resizing and 
detecting the most requested objects in a distributed key-value store
\cite{use-reads-or-not} or notifying backup servers to reclaim the space of 
old logs in order to save new ones \cite{drtm+h}.
Additionally, as already explained in the previous
paragraph, some distributed data structures, like the indexes of key-value stores,
might require multiple indirections inside the index to search for the value of 
the requested key \cite{dist-tree}. If the key-value store uses READ to fetch 
values directly from memory locations, it will have to traverse the index one
node at a time until the location of the value is reached. If a part of 
the index that has to be traversed resides on the same server, then an RPC 
would lead to less round trips than READs, since the servers can traverse 
their own parts locally without the need for new client requests.
Traversing indexes by following pointers in a data structure has also been
called \textit{pointer chasing}, and some key-value stores
resort to RPCs to avoid it, like HERD \cite{herd-kv}.

Distributed systems can also benefit by using both one-sided RDMA operations 
and RPC, to get the best of both worlds. 
For example, Storm \cite{storm-tran}, a transactional dataplane for
remote data structures, uses READs to fetch data at first and only if the 
data reveals that \textit{pointer chasing} is needed, it switches to RPCs.
Thus in case the first try reveals that the remote data structure must be 
searched to find the requested data, switching to RPC 
can reduce the number of round trips, as explained in the previous paragraph.
Another interesting combination of READ and RPC is found in RFP \cite{rfp},
which provides RPC interfaces to access distributed data structures. 
In RFP clients retrieve responses from servers themselves with either
READ or RPC, instead of the servers sending back RPC responses always.
The clients first use READ to fetch the server response 
repeatedly until a configurable threshold of failures. After the threshold is
exceeded, because the server is still processing the request, they use RPC. 
This approach allows the server to save CPU cycles by not sending the response
itself, unless the threshold of retries is exceeded and an RPC is used by
the client.
Moreover, the message size may also determine whether the message should be sent
via RPC or one-sided operations, like in AR-gRPC \cite{ar-grpc}, an RDMA solution 
for gRPC. AR-gRPC sends small messages with a two-sided RDMA based RPC, 
while large messages are split into chunks, which the receiver fetches one by one
with READ. 
This chunking
mechanism avoids blocking the sender, since the receiver fetches the chunks itself. Moreover, chunking also
allows the receiver to consume partial messages, instead of
blocking until the full transfer 
finishes.

There can be also cases where different parts of an application require 
different approaches. For instance, in Octopus \cite{octopus}, a Distributed
Filesystem, metadata is kept private while the rest of the data is shared 
among a cluster. Octopus uses RPC to enable the server to decide whether to 
allow the metadata access or not, while READ and WRITE are used to provide fast 
and direct access to the shared data.

\subsubsection{One-sided RDMA Versus RPC For Remote Locks}

The dilemma of using either one-sided RDMA or RPC is also translated 
into the choice of implementing remote locks with either RDMA atomics
or RPC. Here, a remote lock means that a machine can ask another machine
to lock or unlock access to some protected memory region in that second 
machine.
RDMA supports the atomic operations \textit{fetch-and-add} and
\textit{compare-and-swap}, which are one sided operations, as they 
act on a remote memory location without requiring processing on the other side. 
The remote NIC is the one that guarantees atomic access on a memory location.
Therefore, remote locks can be made by using the RDMA atomics
to change the value of a lock on another machine to indicate
whether some protected data is locked or unlocked. 
One big downside to this approach though is that most RDMA NICs do not guarantee
atomic access between the NIC and the CPU of a server, or between multiple 
NICs on the same server. That means that race conditions can still happen if 
both CPU and NIC or multiple local NICs access the same memory region.
Hence, some distributed systems that require remote locks use either 
RDMA atomics with techniques to avoid the aforementioned race conditions 
or resort to RPCs.

RPCs are more convenient in building remote locking mechanisms than RDMA atomics.
One reason is that the servers that get the locking requests can simply use any
existing locking mechanism offered by the Operating System, without having to 
worry about race conditions between multiple local NICs or the CPU and the NIC.
Another reason is that with RPCs, the servers that receive remote lock requests
can also perform other tasks related to locking, before sending a response.
This can eliminate the necessity to send subsequent requests to the server 
to tell it what to do after the protected area is successfully locked, which
is what would have to happen with RDMA atomics.
For instance, FaSST \cite{fasst-tr} is a distributed transactions system that
operates on a two-phase commit protocol. The coordinator asks the primary servers
to lock their key's headers and send the corresponding values back. 
With one-sided RDMA operations, these two tasks would require two
round trips, one RDMA atomic operation for locking and one to read 
the value. An RPC though performs both tasks with one request.

On the other hand, RDMA atomics, being one-sided operations, can lead to 
less latency and more efficient CPU usage on the remote side than RPCs, 
for remote locking. Distributed systems that use these atomic operations though, have to avoid the aforementioned race conditions that can
occur.
For instance, in the Shift \cite{sift-cons} distributed consensus system, during leader election,
the servers attempt to write election messages to each other's memory with the 
RDMA \textit{compare-and-swap} (CAS) atomic operation. 
Only one machine will succeed with CAS on another machine, so the one that has
written successfully to a majority of machines becomes the new coordinator.
Due to different NICs trying to access the same memory location atomically, this method does not cause race conditions with the CPU or
multiple local NICs.
Another example is the DrTM \cite{drtm-tr} distributed transactions
system, 
that combines Hardware Transactional Memory (HTM) with RDMA atomic operations.
DrTM first locks and fetches remote records, and afterwards it executes the 
transaction locally using HTM. Remote locks are acquired and released with 
RDMA CAS, but the local access inside an HTM transaction will only check the 
state of the locks, without changing them, which can work correctly because 
of the cache coherence of RDMA memory. Thereby an HTM transaction will be able
to see if the local records that it has read or written during the transaction
have been locked by another machine. If that is the case it aborts, otherwise
it continues the transaction.

Similarly, in the Octopus \cite{octopus} distributed filesystem, distributed 
transactions are build on top of both GCC (GNU
Compiler Collection) and RDMA atomics. 
Locks are acquired and released locally through the GCC CAS operation, while 
RDMA CAS only releases remote locks. This locking and unlocking mechanism
prevents race conditions between CPU and NIC in Octopus.

\subsection{Choosing RPC Frameworks Or Custom RPC Protocols}

The last of the design choices presented here for efficient RDMA usage, 
is whether to build a custom RPC protocol with the RDMA operations or use 
an existing RPC framework that uses RDMA. Although the latter option is 
more convenient, building a custom RPC protocol for a distributed
system can enable optimizations specific to that system, which are not
possible with a general purpose RPC framework. The rest of this section 
explores what does each of the two options offer in more detail, as well
as how they affect the communications performance of a distributed system.

RDMA RPC frameworks can speed up the system design and development 
process. By using such a framework there is no need to go through all 
the design choices for efficient RDMA usage presented so far, or write 
code to implement RPCs over RDMA, since many features are provided 
"out of the box". 
For example, the ScaleRPC \cite{scale-rpc} and
DaRPC \cite{da-rpc} frameworks that offer RPC over
RDMA, deal with the scalability issues of RDMA, as described in section
\ref{rdma_scalability}. 
Additionally, the eRPC \cite{e-rpc} framework offers RPCs over RDMA's
Unreliable Datagram connection, to avoid scalability issues with Reliable and
Unreliable RDMA connections.
Furthermore, the Mercury \cite{mercury} RDMA RPC framework also provides a
way to deal with bulk data transfers, eliminating the need from users of the 
framework to do so themselves.
Mercury can split bulk data transfers to smaller chunks and send a chunk at 
a time, enabling the server to process the data as it flows and not having
to pay the latency cost of waiting until the whole transfer is complete to
start processing it. This can also deal with large messages that don't 
fit into a memory buffer at the server as a whole.
So for distributed systems that are under development, 
an RPC framework can greatly reduce the effort to achieve good 
performance with RDMA as the base for RPC.

Although RPC frameworks are convenient by providing features "out of the box",
they are general purpose solutions, which lack the flexibility of making 
design choices that fit to a specific system's needs. Systems designers that
desire this flexibility to apply customized optimizations for their system 
will have to built their own RPC protocol, which comes with the tradeoff of 
increased complexity. For example, the designers will have to choose an RDMA connection
method and deal with any scalability issues that this connection method 
incurs. 
Moreover, they will have to select with which RDMA operations to implement RPCs,
since it's possible to use either one-sided or two-sided operations.
Regardless of the choice, to build a request-response protocol like RPC, there
must be a way for the receiver to be notified about new messages.

There are different ways to detect new messages, depending on
which RDMA operations are used for RPCs.
With two-sided SEND/RECV, notifications about new messages 
can be retrieved from Completion Queues. 
With one-sided WRITE on the other hand, 
there are two possible options.
The first is to use an \textit{immediate} 
value, which is sent along with the data to
the remote side and is passed as a notification to the receiver.
The second option for message detection with WRITE is to use a polling mechanism. This technique
can achieve less latency than the previous option, 
by avoiding the
notification part, which requires a DMA transfer from the NIC to a Completion
Queue in main memory. With polling, instead of waiting for a notification to 
be sent by the NIC, the receiver directly looks into the memory location where
messages are written to detect new ones.
Polling can cause high CPU usage though, so in \cite{hydradb-kv} and 
\cite{derecho}, \textit{sleep} functions are proposed in order
to have the servers sleep after a period of inactivity. 
Furthermore, it also possible to combine two-sided and one-sided operations
for an RPC protocol like HERD \cite{herd-kv} does, which allows HERD's server 
to handle multiple client connections in a scalable manner, as described in
\ref{rdma_connections}.

In conclusion, RDMA offers high performance communications but at the cost
of complex design decisions. Since a high level abstraction like sockets is
not provided by default, users of RDMA have to either build their custom 
protocols or use existing RDMA frameworks. While the former option offers more 
freedom of design choices and the ability to make customized optimizations
according to system requirements, the latter option makes both design and 
development easier. Moreover, building a custom solution requires spending
time to study research papers in order to learn about what has been tried
before and how can certain design choices affect the performance of RDMA 
communications. This task would also involve learning low level optimizations,
like how can networking requests to the NIC be shrinked or batched to 
maximize performance \cite{rdma-design}, thus requiring a higher level of 
expertise on the RDMA technology for developers than using simpler, socket-like
APIs. 
Nevertheless, it is up to the users of RDMA to decide whether they wish a 
more convenient solution like an existing framework or a custom solution in order
to achieve the maximum performance for their specific system, since both options
are available.

\section{Programmable NICs}

Another hardware-based approach to high performance distributed 
system communications is to use programmable NICs, also called 
SmartNICs.
SmartNICs typically contain multicore processors and onboard memory,
enabling generic computations on the NIC, although their resources
are limited.
Because of their programmability and their on-board resources, it
is possible to offload networking functionality on them.
Additionally, it's possible to build networking protocols on them 
and bypass the OS kernel for networking operations, similarly to
RDMA. Hence this is categorized as a kernel bypass approach too.
Moreover, the \textit{programmable} feature allows developers to change
that functionality any time without requiring new equipment. 
Therefore, programmable NICs are another option for 
achieving greater communications performance
than the standard OS kernel network stacks.

Based on the idea of offloading networking operations to programmable NICs,
the Tonic paper \cite{tonic-nic} proposes a hardware design to enable
programming transport logic for network protocols run on those NICs. 
\textit{Transport logic} refers to 
choosing which data segments to transfer (data delivery) and when
(congestion control). Programmability allows changing the 
algorithms for data transfer and congestion control on the NIC, 
according to which features are desired by a system. 
It is therefore a more flexible solution than offloading specific network
protocols like RDMA or TCP on non-programmable hardware.
Given the mismatch of certain
TCP features with the needs of RPCs presented in \ref{ineff_tcp},
Tonic would also be useful in implementing network protocols that match
these needs and increase efficiency.
The hardware design proposal in Tonic \cite{tonic-nic} could 
possibly replace 
hard-wired RDMA and TCP protocols on NICs in the future, as it
provides more freedom in designing custom network protocols,
while also saving CPU cycles by offloading operations on the NICs.

Besides using SmartNICs for building network protocols that bypass the OS kernel, 
Microsoft's Azure Cloud has also used SmartNICs along with SR-IOV
to bypass their hypervisor's network stack, in order to 
reduce latency and improve throughput in networking operations.
In a traditional device sharing model, like a public cloud,
all network IO to and from a physical device is performed by
the host software partition of the hypervisor.
That means that packets that are received are processed
by the hypervisor before reaching the Virtual Machine that they
were intended for. Similarly, when sending packets, those packets
are processed in the hypervisor before being transmitted over the 
network.
This additional processing by the hypervisor degrades performance
in communications, by increasing latency, latency variability and CPU utilization, 
while lowering throughput.
The SR-IOV technology can be used to improve performance by bypassing the 
hypervisor's networking stack.
It does so by creating Virtual Functions (VF) that represent a physical networking 
device and by assigning each Virtual Machine (VM) its own Virtual Function.
Then networking traffic can flow directly between a VF and a VM, without being
processed in the hypervisor.
This bypass though also includes bypassing any SDN
policies that were implemented in the hypervisor.
To still be able to enforce such policies when using SR-IOV, 
while maintaining programmability and flexibility on the design of 
virtual networking features, the enforcement of the policies was offloaded to
SmartNICs that use FPGAs for programmability.
This approach led to less CPU usage, latency and latency variability
and achieving line rates even on a single flow for VMs 
with 32 Gbps network capacity.

Besides their advantages, programmable NICs also come with certain
tradeoffs. 
First of all, the availability or ownership of such hardware 
is necessary in order to reap its benefits. 
Second, developers specialized on programming SmartNICs are needed to
build network protocols on these NICs, and provide interfaces for
applications to use these protocols.
A third tradeoff is the limited resources 
(computing/memory) 
on the hardware compared to a server, that put a barrier into
how much functionality can the NICs handle. 
Nevertheless, offloading networking operations on SmartNICs
can lead to great performance gains, since CPU cycles are saved and
networking is performed by programmed hardware, instead of the OS kernel 
software network stack.

There have also been proposals to offload
entire distributed systems or parts of their functionality
to programmable hardware like NICs, FPGAs and network switches,
along with their network stacks \cite{consensus-in-a-box},
\cite{canns-consensus}, \cite{netchain-switches}, \cite{ipipe-smart-nics}.
Although examining the offloading of computational operations to such 
hardware is out of scope of this survey, which focuses on communications only, 
there is an interesting aspect on those proposals;
co-locating the distributed system logic with the networking logic on the 
same chip can reduce the latency of communications, since the processing
and networking hardware components are closer to each other than in 
a traditional motherboard system. 
This idea has also led to research on using SoCs
(System-on-a-Chip) to
accelerate distributed system communications, as will be presented next.

\section{System-on-a-Chip}

Besides RDMA and programmable NICs, research has also investigated the use of
System-on-a-Chip in implementing network protocols in hardware, that enable 
bypassing the OS kernel's network stack.
System-on-a-Chip (SoC) is an integrated circuit that combines multiple computing 
resources like CPU cores, memory, IO ports and even Network Interfaces 
on the same chip, leading to faster data transfers between them.
In contrast, traditional PC 
architectures separate these components by functionality and connect 
them via a central motherboard.
Therefore, building networking protocols on SoCs that integrate CPU, memory, NICs
and routers makes it possible to achieve even lower latency than RDMA and programmable
NICs.
Hence, this option is classified as a hardware-based, kernel bypass approach.

In soNUMA \cite{so-numa} and \cite{ni-arch}, hardware designs are proposed
to bring RDMA-like protocols to SoCs that integrate CPU, memory, a NIC and a router. 
Because of the integration of the NIC to the processor's cache 
hierarchy, the latency of networking operations is reduced greatly, 
compared to using a PCIe attached network card in 
a traditional motherboard-based system.
This enables these SoC designs to achieve lower latency than using RDMA network
cards/adapters or programmable NICs.
But like the two latter approaches, the networking operations are still offloaded 
to specialized hardware, bypassing the OS kernel.

Another interesting way to leverage the faster CPU to NIC interactions that
SoCs can provide, is implementing accurate real-time load balancing mechanisms 
between cores that serve RPC requests, as proposed in RPCValet \cite{rpcvalet}. 
The tight integration of NIC and CPU on a SoC can
enable the NIC to monitor real-time networking traffic load on the CPU cores 
and implement load balancing policies on the distribution of this traffic to 
the cores.
The difference with other load balancing mechanisms of NICs distributing
network traffic to CPU cores like Receive Side Scaling, is that these use 
static rules for load distribution that do not respond to real time events.
To enable real-time load balancing, RPCValet introduces the 
new operations \textit{send} and \textit{replenish}, which, like RDMA 
operations, are based on the VIA interface \cite{via}.
The \textit{send} operation sends a message to a remote machine and specifies
which buffer will store the message in the receiver, while the
\textit{replenish} is sent back to a sender to notify it that the request has
been processed and the receive buffer has an empty slot to accept a new message.
The \textit{replenish} 
operation is important for the NIC to know when a request has been processed
and therefore update the number of outstanding requests on a core, which 
is the unit of measurement of per-core load. 
The integration of CPU and NIC on a SoC allows \textit{replenish} operations
in RPCValet to reach the NIC fast enough to keep real-time load information,
which would not be that accurate in a slower motherboard-connected 
hardware system.

Although the integration of networking and processing components in SoCs 
can accelerate
distributed system communications and enable more accurate load balancing,
this approach requires servers that run on SoCs.
While SoCs are commonly
found in embedded systems like smartphones or tablets because of 
their small size and reduced power consumption, the AWS cloud provider
recently started to offer its customers the option to use server instances 
that run on ARM processors.
These ARM processors are SoCs designed to minimize power consumption
and heat on data center servers, which can lead to
further adoption of SoC-based servers
in clouds. This adoption can in turn motivate further 
research, development and usage of SoCs that integrate networking and CPU 
components and enable kernel bypass, in order
to decrease the latency of communications in data centers.

\section{Software-based Approaches}

Leveraging specialized networking hardware to improve distributed system 
communications, can create a tight coupling of the system's 
communications layer design with the hardware capabilities.
As it was shown with RDMA, using this hardware technology efficiently
requires making design choices that depend on the available features of 
the technology.
This can also happen when using networking protocols built for programmable NICs
or SoCs.
Such design decisions though can become outdated, as 
new hardware may support new features that can redetermine which
choices are more beneficial for a system.
Therefore, research has also been conducted on improving the performance
of distributed system communications through software, in order to avoid
relying on specialized networking equipment. 
The software-based approaches presented here are
optimized socket implementations and 
RPC frameworks, as well as user space networking techniques,
which include packet processing frameworks, user space TCP stacks and 
data plane operating systems.

\subsection{Optimized Socket Implementations}

Besides kernel bypassing techniques, there are also proposals to
optimize the existing communication utilities provided by operating system
kernels, by offering more efficient socket primitives. These solutions 
can improve the performance of communications without requiring 
specialized networking equipment. 
Moreover, they offer a more familiar and abstract interface to application
developers, than using low level primitives, as with RDMA or any similar 
protocols implemented in hardware.
Their main drawback though is that since the kernel is not bypassed,
they do not solve all of the overheads that sockets and the kernel
network stacks impose during data exchanges.

The purpose of optimized socket implementations proposed in literature is
to deal with the performance overheads of sockets explained in 
section \ref{ineff_socks}. As a reminder, the overheads examined here are 
induced by the internal socket mechanism that allows only one thread to accept 
a new connection at a time and the treatment of sockets as file system objects,
thus inheriting file system related overheads.
To solve the issue of only one thread accepting a connection, MegaPipe 
\cite{megapipe} and Affinity-Accept \cite{affinity-accept} propose the 
replacement of a socket's internal synchronized \textit{accept queue} 
that accepts new connections, with one \textit{accept queue} for each thread 
that handles connections on that socket. 
Thus a socket can be shared among cores without the
need for synchronization.
A similar effect can be achieved by the Linux kernel though, by allowing multiple
sockets run in different threads to listen on the same socket address through
\texttt{SO\_REUSEPORT},
while the kernel handles load balancing the connections to these sockets.
Affinity-Accept also aims in maintaining the CPU core affinity for connections,
by accepting a connection on the core that was already assigned by the NIC to 
process the packets for that connection. 
This method can lead in better performance, as data related to the connection
remains on the same cache. 
Moreover, due to being treated like file system objects,
sockets are associated with globally-visible file system data, which makes them
susceptible to system-wide synchronization and also require a time-costly 
file descriptor allocation. To solve these problems as well, MegaPipe provides
its own socket implementation that is not related to the file system.

Techniques that optimize sockets 
offer a performance increase for distributed systems communications, 
but only solve a small subset of the issues in \ref{ineff}. System call 
overheads, context switches and redundant copies between user space and kernel 
space, remain.
Furthermore, the kernel's network stack is still used, which means that packet
processing remains slow as well. 
Therefore, such techniques can not offer the same performance optimizations
as kernel bypass approaches.

\subsection{Optimized RPC Frameworks
\label{rpc-frameworks}}

Looking at a higher level than sockets, there has also been effort to 
improve the efficiency of RPC frameworks, with the goal to achieve
low latency and high throughput for distributed system communications.
Since TCP can hurt the performance of RPCs, as mentioned in \ref{ineff_tcp},
certain RPC frameworks switch to UDP based communications
to build more efficient RPC protocols.
Like optimizing sockets, optimizing RPC frameworks is not a kernel bypass approach,
but can aid in increasing the performance of RPCs.
Nevertheless, it will be shown that it's possible to combine optimizations
on RPCs with \textit{user space networking} techniques, 
which bypass the OS kernel for networking operations, 
and achieve performance close to RDMA.

The mismatch of some TCP assumptions with RPC semantics has led to the creation
of RPC frameworks that operate on UDP, like eRPC \cite{e-rpc} and 
R2P2 \cite{r2p2-rpc},
in order to meet the performance requirements of large scale applications.
As explained in R2P2
\cite{r2p2-rpc}, RPC requests and responses usually fit into a
single packet and since they are semantically independent, there is no need
for TCP to order these packets.
Moreover, to amortize the cost of setting up TCP flows, RPCs are often layered
on top of a persistent TCP connection,
which leads to ordering packets even if they originate
from different RPCs. This connection sharing can also lead to head-of-line
blocking of independent RPC packets.
UDP on the other hand does not enforce any packet ordering, allowing RPC
semantics alone to distinguish the requests from the responses. Furthermore,
unlike TCP, UDP does not setup and maintain connection information, 
leading to greater scalability than TCP with a large number of clients and
servers, while eliminating the need to share persistent connections.
R2P2 uses UDP to also enable the implementation of various 
load balancing policies for processing RPCs.
More specifically, clients forward their requests to an intermediate router,
which selects the server that will process the request and send a
response, based on a
load balancing policy. With UDP there is no need to setup a new point-to-point 
connection between a client and the server that is chosen to handle the 
RPC every time, thus saving time, memory and computing resources.
Apart from these advantages though, a UDP based RPC would have to implement
custom methods of handling packet losses, congestion and flow control,
unlike TCP which provides this "out of the box".

The eRPC framework \cite{e-rpc} has also proven that it's possible to 
get close to RDMA performance for RPCs, with a general purpose RPC 
framework. 
Its goal is to achieve high performance RPC communications, even without
specialized networking equipment.
eRPC supports both UDP for lossy Ethernet networks 
and Unreliable Datagrams when Infiniband is available. Unreliable transports
help eRPC solve the mismatch of TCP and the TCP-like Reliable Connections of RDMA with RPCs, regardless of networking infrastructure. 
eRPC also uses three additional optimizations, to be able to 
reach RDMA performance, even when Infiniband is not available.
The optimizations are
writing code optimized for the \textit{common case},
restricting each flow to at most one Bandwidth-Delay
Product (BDP) of outstanding data, 
and using \textit{user space networking} techniques.
In eRPC, the \textit{common case} refers to the fact that usually 
the network is uncongested, the messages are small and RPC handlers
are short. Therefore eRPC's code is optimized to execute fast when these
conditions hold.
Additionally, restricting the flow of messages to BDP can allow eRPC to 
reach a throughput equal to the network link's data transfer rate 
without dropping packets, since network switches have large enough packet
buffers to support BDP tranfer rates without packet drops.
The third and very important optimization is to use \textit{user space
networking} techniques, which actually bypass the OS kernel 
for networking operations without specialized hardware, but with software.
More specifically,
zero-copy transfers are achieved by dedicating DMA-capable buffers for messages, 
similarly to what RDMA does, which are owned and accessible directly by 
the applications in the user space. Thus,
intermediate copies of these buffers are avoided, and both applications
and NIC can access them without requiring the kernel's involvement.
The performance analysis in eRPC shows that all the above features allow it to
perform almost as well as specialized RDMA solutions, even by being a
general purpose RPC framework.

Therefore, besides optimized socket implementations, optimizing
RPC frameworks is another option to improve the performance of
distributed system communications through software.
One advantage of these frameworks over the optimized sockets is the provisioning
of an RPC API to developers, that they would otherwise have to build themselves.
A drawback is that optimizing RPC frameworks by making RPC
implementations more efficient does not solve the overheads 
with sockets and the kernel network stacks.
Nevertheless, combining optimizations on RPC frameworks with
kernel bypassing techniques like eRPC does, 
is a great alternative to relying
on specialized networking equipment to enable the bypass. 
This software-based kernel bypass is not a new concept though, as it has been
a research topic for years in the scope of \textit{user space networking}, 
which will be presented in the following section.

\subsection{User space Networking
\label{us_net}}

User space networking means managing networking operations in the 
operating system's user space, instead of the kernel space. 
The kernel's network stack is bypassed and networking operations
use a network stack running in the user space instead.
Packets are directly stored and processed in
buffers shared between the NIC and the applications,  
without intermediate data copies happening on the kernel (zero-copy transfers),
achieving faster communications.
Additionally, it is considered more convenient for developers to 
build and optimize network stacks in the user space than attempting 
the same with complex and large kernel code.
Furthermore, the lack of tight coupling to specific networking equipment
like RDMA, makes user space networking a more widely available option.

A very important feature of user space networking that affects the performance
of communications its ability to separate networking operations to
\textit{control} and \textit{data} planes. 
To avoid confusion with the common use of these terms in network switches, 
in this text these two planes refer to operations that take
place in an operating system (OS) to transfer and receive messages.
They are practically the same as the control and data "paths" for RDMA 
presented in \ref{rmda_net}. 
The \textit{control plane} is comprised of operations related to the 
allocation and management of I/O resources, like pre-allocating message 
buffers for the communications or implementing security checks on access 
requests. The \textit{data plane} refers to the actual data transfer 
operations.
In a traditional OS, these two planes run both in the kernel space, but in 
user space networking, the data plane runs in the user space.
Separating these planes allows the more time-consuming \textit{control plane} 
operations to take place only when needed, instead of repeating them on every 
data transfer, reducing the time cost of communications. 
Moreover, if the data plane fails it can be restarted 
independently, whereas in a kernel network stack the \textit{control plane}
would fail as well, leading to a slower restart of communications.

Due to achieving low latency and high throughput comparable to hardware-based
approaches for kernel bypass, user space networking is a topic that has caused
a lot of interest in the past decade.
During that time, this interest led to 
the development of various tools for packet processing, entire TCP network 
stacks run in the user space and even \textit{data plane operating systems},
that combine user space networking with the OS kernel, to provide fast but 
also secure communications. 
The survey proceeds with providing more details on these user space
networking implementations.

\subsubsection{User space Packet Processing Tools}

User space packet processing tools
offer applications the ability to store and manage packets
directly in the user space for zero-copy transfers,
while they also optimize packet processing. For instance, the DPDK 
\cite{dpdk} framework processes each packet with a \textit{run to
completion} 
model. This model completes the application and network protocol processing
for a packet before moving to the next one. The OS kernel on the other
hand allows interrupts between processing stages, which requires 
intermediate buffering of data to proceed between those stages. This allows
the kernel to execute scheduling mechanisms for other tasks but slows down 
packet processing. With \textit{run to completion} DPDK eliminates the 
need for intermediate buffering and reduces the latency of processing a 
packet. The netmap framework \cite{netmap} also offers the ability to 
send and receive multiple packets at once, amortizing the cost of per
packet management.

User space packet processing tools also implement the
separation of operations to control and data planes as described in
\ref{us_net}. 
By assigning the slower control plane to the kernel and minimizing the 
time that the control plane is involved in the communications, 
they can achieve throughput that utilizes the available 
link's bandwidth as much as possible. 
For example the DPDK \cite{dpdk} and netmap \cite{netmap} user space 
packet processing frameworks pre-allocate packet buffers through the kernel
before commencing the communications to avoid repeating this expensive 
operation on each transfer. 
In netmap, the kernel is actually called to request the NIC to send packets or
to ask the NIC if any packets were received, 
but any packet processing still runs on the 
user space. The kernel can also perform these operations 
for multiple packets at once, amortizing the cost of the system calls.
All these techniques that reduce the time that the operating system kernel 
is involved in communications, provide this saved time to data plane 
operations that perform the data transfers, thus achieving less latency and
higher throughput.

It should be mentioned here that besides bypassing the kernel to provide
more efficient packet processing stacks in the user space, 
it has also been attempted to improve the packet processing performance 
from within the OS kernel in XDP \cite{xdp}.
XPD's purpose is to reach performance close to user space packet processing 
tools, but without kernel bypass.
One reason for keeping packet processing inside the kernel is to be able to
use functionality provided by operating system network stacks without having to 
reimplement everything as user space packet processing tools have to do.
Another reason is that direct hardware access from user space applications poses
security risks.
To enable more efficient packet processing inside the kernel, 
XDP uses eBPF to run programs that process packets before the kernel does so.
These programs can completely avoid the kernel network stack or only use 
certain features from the stack that they require, but their code is still 
run within the kernel in either case.
Although offering a great improvement in performance compared to 
the kernel's network stack, 
this approach does not make possible all the optimizations that 
user space packet processing offers, as some overheads remain.
This is why XDP is not as fast as user space packet processing solutions
and at a packet drop benchmark in the XDP paper \cite{xdp} 
with only one core, XDP managed to process
24 million packets per second (Mpps), while DPDK 43.5 Mpps. The authors
claim that future work can decrease the performance gap to user space
packet processing tools even further.

Although user space packet processing tools achieve higher
performance in communications than traditional OS kernel network stacks, 
they also have some drawbacks. 
One of them is that they are not inter-operable with existing kernel tools 
for network management, since kernel tools do not interface with the user 
space network stack. This might be inconvenient for people used to these 
kernel tools for network management.
Moreover, because they were created for applications like firewalls, 
routers and traffic monitors, they lack a higher level abstraction like 
sockets, making it harder for developers to write code for communications.
Nevertheless, they have been used as building blocks for creating more 
efficient network stacks in the user space, which come along with interfaces 
for developers similar to sockets. One such example is user space TCP stacks,
which will be discussed next.

\subsubsection{User Space TCP Stacks}

User space TCP stacks, as their name suggests, implement the TCP 
network stack into user space. The difference from user space 
packet processing tools is that they use such tools to build more efficient
TCP stack implementations than the ones provided by operating system 
kernels. 
They also offer socket-like APIs to developers, thus requiring less
code for communications than packet processing tools.

To provide faster packet processing for TCP connections, these stacks 
have to deal with the problems of in-kernel TCP stack implementations 
outlined in section \ref{ineff_kern}. These problems consisted of delays 
related to inefficient cache usage, locks and per-packet memory allocations 
and de-allocations.
Furthermore, the TCP state 
machine's code is run a monolithic code block that does not allow 
opportunities for batching and prefetching, while it consists
of many branches, increasing the instruction cache memory footprint and causing
branch mispredictions \cite{tas-tcp}.

By creating custom TCP stacks in the user space, these problems can be 
solved. For instance, to use CPU caches more efficiently, the 
mTCP\cite{mtcp} user space TCP stack creates one TCP thread for each 
application thread, and collocates the application threads with their
corresponding TCP threads on the same core, 
preserving affinity and reusing cached data more often.
Moreover, mTCP minimizes the size of data structures associated with
packets and connections and aligns them to CPU cache lines to prevent false
sharing.
By keeping these data structures local to the TCP threads, locks are also 
avoided. Furthermore, to avoid frequent memory allocations and de-allocations
mTCP creates a memory pool per core and also utilizes huge pages, to reduce 
misses on the Translation Lookaside Buffer.
Finally, mTCP uses a packet processing engine
that transmits and receives packets in batches, amortizing the costs of
expensive operations over PCIe during the communications. 

The TAS \cite{tas-tcp} TCP stack has also shown that it is possible to 
further accelerate TCP packet processing, by separating the monolithic 
OS kernel TCP code to common and uncommon cases. TAS's authors claim 
that within a datacenter, packets are usually delivered reliably and in order,
not fragmented and without time-outs firing. This is considered
the common case, while 
the uncommon case is when the opposite happens for the packets.
TAS splits the TCP \textit{data} plane (section \ref{us_net}) even further, by
handling the common case on a \textit{fast path} and 
the uncommon case along with other time-expensive operations on a 
\textit{slow path}.
The fast path processes headers, sends acknowledgements, segments 
the payload and enforces network congestion policies. The slow 
path implements policy decisions and management mechanisms that 
do not have a constant overhead per packet or are too expensive, in terms
of time, to perform in the fast path.
For example, it retrieves per connection feedback from the fast path and
then runs a congestion control algorithm to figure out a new flow send rate
or TCP window size. It also handles connection setup/tear down and detects
retransmission time-outs, in which case it can notify the fast path to 
retransmit any packets that timed-out.
The separation of the TCP monolithic code to fast and slow paths reduces the
time spent on packet processing for the common case, which means executing
less code per packet most of the time, achieving lower latency and higher 
throughput.

In conclusion, user space networking can greatly reduce packet processing 
time costs, allowing network stacks to operate with a throughput very close
to network link rates.
That means that distributed systems are able to fully utilize high speed
network links, thus minimizing communication overheads. User space 
networking techniques though have been criticized for lacking 
sufficient protection, since untrusted user space 
applications might corrupt the network stack, even through bugs and crashes
\cite{ix}. This has led to the development of \textit{data plane operating
systems}, 
that combine user space networking stacks with the protection mechanisms 
of an operating system and are the next topic of this survey.

\subsubsection{Data plane Operating Systems}

Data plane Operating Systems implement user space network stacks, while they 
also employ the Operating System kernel to manage communications.
Like user space packet processing tools and network stacks, 
they separate 
networking operations to \textit{control} and \textit{data} planes, 
assigning the former to the kernel and the latter to the user space 
network stacks. But unlike the previous two user space networking approaches,
they rely on virtualization technologies to isolate applications and protect
access to resources more efficiently.

Allowing applications to manage packets and network stacks 
directly through interfaces, leaves the use space network stacks liable to 
malicious or erroneous application behaviour that can affect other 
applications on the same machine \cite{ix}. 
Data plane operating systems solve this issue 
by applying an extra layer of protection through hardware virtualization 
and kernel involvement. With virtualization, applications are assigned
virtual devices and are isolated from other each other. Moreover, packets
can be delivered to the virtual interfaces without the need for a copy
inside the kernel. The kernel on the other hand is still responsible for
protecting access to resources. For example, in the IX data plane 
operating system \cite{ix}, 
data planes are run as application-specific operating systems through
virtualization, and each application is assigned its own data plane.
The data planes are run in a different privilege ring than user space 
applications, thus having access to hardware resources while not allowing
the applications to access data plane memory besides message 
buffers. The kernel also isolates data planes from each other, thus 
there is no way for them to access each other's memory; an 
application can't corrupt the network stack of another application.
Similarly to IX, the Arrakis \cite{arrakis} data plane OS uses the 
\textit{Single-Root I/O Virtualization (SR-IOV)} technology to 
create isolated virtual network cards for applications, 
managed by the kernel (control plane). 

Besides user space networking features, data plane operating 
systems have been applying various optimizations to further 
increase communications performance.
For instance, IX \cite{ix} applies batching when possible to various stages 
of the network stack in order to amortize the costs of packet processing. 
It only uses batches when they are already available though, without waiting
for a batch to complete before proceeding, thus not sacrificing latency.
Another interesting feature found in the ZYGOS \cite{zygos} data plane OS,
is work stealing of connections between cores, which enables better load
balancing on CPU cores and avoids \textit{head-of-line blocking} in situations
with a large number of connections. This approach introduces intermediate
buffering though and might hurt the performance when the tasks assigned to the
cores are very small. 

It is also worth mentioning the Snap \cite{snap} microkernel-like approach here,
which is very similar to data plane OSes, but also offers RDMA-like 
one-sided operations. Snap functions like a microkernel, putting both 
control and data plane in the user space and running as a process. Unlike
microkernels though, it does not require the entire system to adopt the 
microkernel design.
Besides implementing user space networking similarly to data plane operating
systems, Snap's most interesting feature is that 
it also supports RDMA-like one-sided operations.
These operations do not involve any
remote application thread and they execute to completion inside Snap's engine, 
saving CPU cycles. 

The application isolation and additional optimizations of data plane operating
systems and similar microkernel designs, make them a more attractive alternative
for user space networking than using packet processing tools or network stacks
directly.
Virtualization along with the kernel's access protection mechanisms provide
an extra layer of security and system resilience against untrusted
applications. 
This is especially important for deploying user space networking solutions
in production environments, in which multiple applications might share the
same networking devices.

\section{Discussion}

The main research question of this survey is :

\textit{"What has been proposed in research literature between 2010-2020 to
overcome communication overheads imposed by  sockets, kernel network stacks and TCP on
RPCs?"}

The survey answered this question by presenting hardware and
software-based approaches on overcoming the aforementioned overheads,
found in research literature published between 2010-2020.

From all the approaches presented, 
hardware and software-based kernel bypass techniques
have gained a lot of attention in research over the last 10 years, 
since they offer very low latency and high throughput in communications.
Their high performance is the result of dealing with the overheads of 
OS-provided network stacks and socket implementations.
Moreover, they can also offer an alternative to using TCP for RPCs, 
by enabling developers to build custom protocols optimized for RPCs or use
unreliable transports, as proposed in section \ref{rpc-frameworks}.
Thus they provide a more complete solution to the problems with 
traditional communication methods presented in section \ref{ineff},
than socket and RPC framework optimizations alone.
Furthermore, besides research and use in privately owned infrastructure, 
kernel bypass solutions have appeared in public clouds too.
For example, AWS offers the Elastic Fabric Adapter, which allows 
applications to bypass the kernel's
network stack and communicate directly with the network interface hardware,
through a software middleware.
Moreover, RDMA has also been incorporated in data centers of certain cloud 
providers (e.g. Microsoft Azure, Oracle Cloud).
The incorporation of kernel bypass solutions in cloud data centers
proves their ability to offer high communications performance in large
scale environments and
can motivate further research on optimizing these technologies or 
adding new features in the future. 

The hardware and software kernel bypass approaches 
that were presented in this survey
differ in their reliance on hardware equipment, 
which has comes with certain tradeoffs.
One of the tradeoffs of hardware-based kernel bypass is that 
acquiring specialized hardware like
programmable NICs, SoCs and RDMA-capable network cards
leads to extra costs.
Software-based RDMA (Soft-RoCE and SoftiWARP) doesn't achieve the high
performance of RDMA-capable hardware and hence it's less suitable 
for data centers.
Moreover, the above hardware-based approaches
lack a high level interface like sockets, which makes them harder to
use by application developers.
Another important fact is that distributed system design can become 
more complex, when the networking hardware capabilities require making certain decisions when designing the communications part of 
the system. 
User space networking techniques
on the other hand can provide communications performance comparable to 
the hardware kernel bypass approaches,
without relying on specialized networking
equipment to achieve high performance. 
They might only require certain hardware features
like support for virtualization.
The downside to user space networking is that it doesn't benefit from
hardware offloading, which can save CPU cycles and lead to greater 
communications performance, as networking functionality is built directly 
in hardware instead of software. 

Therefore, the answer to the second research question 
\textit{"Which one is better for distributed systems communications,
hardware or software-based approaches?"} is :

\textit{
\textbf{Both hardware and software-based approaches can achieve very high
communications performance for distributed systems, through kernel bypass.
Moreover, 
choosing between specialized hardware or user space networking for kernel bypass 
is a matter of
which of their tradeoffs are more tolerable to distributed system designers and developers,
rather than which technology is better}}.

\section{Limitations}

This part describes limitations of the survey.
One of those limitations is that the overheads in communications
in section \ref{ineff} refer to the Linux kernel, which is open source. 
It is not possible to examine the code of other closed source operating 
systems that run on servers (e.g. Windows), which is why papers
discovered during the searching methodology (section \ref{method})
describe inefficiencies in distributed system communications on Linux. 
Moreover, as already described in \ref{method}, the papers selected 
focused on technologies for servers only, although there have been some solutions
for switches, routers or FPGAs that can also speed up distributed system
communications. 
Looking into other hardware besides servers for networking or any software equivalents, like software routers,
would make the topic of this survey too broad. 
Hence, the decision was made to focus only on technologies and techniques that
can be applied to servers, which might be considered as another limitation.
Similarly, GPUs have also been proposed to accelerate packet processing 
computations with software, often for software router implementations.
Therefore, since this approach focuses mainly on accelerating computations
during packet processing than on solving inefficiencies with the network stack,
and it's application was proposed for software
routers in most of the papers discovered during search, 
it was also not included here.
Nevertheless, approaches that were excluded from the survey are mentioned here, in order for
interested readers to be aware of them, in case they wish to look into them.

\section{Conclusions}

The inefficiencies in operating system network stacks 
and sockets, as well as the degradation of performance caused to 
RPCs from certain TCP features, 
have motivated research on improvements and 
alternatives on distributed system communications for years. 
This survey presented solutions to these overheads proposed in literature
during the recent decade at the time of writing, 2010-2020.
These advancements have been categorized to hardware-based and 
software-based approaches, according to their reliance on 
specialized hardware to achieve high performance.
The hardware-based approaches were RDMA, programmable NICs
and SoC, while the 
software-based ones where optimizations on sockets and RPC frameworks,
and user space networking techniques.
These approaches have been further classified into \textit{kernel bypass} or 
\textit{not kernel bypass} solutions, according to whether they bypass 
the OS kernel's network stack or not.
The kernel bypass solutions, which were RDMA, networking protocols on 
programmable NICs and SoCs, as well as user space networking,
offer the
maximum benefit by avoiding the kernel's slow network stack and providing
their own alternative implementations, that achieve low latency 
and high throughput. 
Moreover, they can offer alternative ways to build
RPC protocols instead of using TCP, which can improve RPC performance as well.
Both hardware and
software-based approaches for kernel bypass have both 
benefits and tradeoffs.
When choosing which type of kernel bypass approach is better for 
a distributed system, the distributed system designers have to 
decide which of the benefits and tradeoffs are
more preferable to them.

\bibliographystyle{plain}
\bibliography{references}

\begin{thebibliography}{10}

\bibitem{tonic-nic}
Mina~Tahmasbi Arashloo, Alexey Lavrov, Manya Ghobadi, Jennifer Rexford, David
  Walker, and David Wentzlaff.
\newblock Enabling programmable transport protocols in high-speed nics.
\newblock In {\em 17th $\{$USENIX$\}$ Symposium on Networked Systems Design and
  Implementation ($\{$NSDI$\}$ 20)}, pages 93--109, 2020.

\bibitem{derecho}
Jonathan Behrens, Ken Birman, Sagar Jha, Matthew Milano, Edward Tremel, Eugene
  Bagdasaryan, Theo Gkountouvas, Weijia Song, and Robbert Van~Renesse.
\newblock Derecho: Group communication at the speed of light.
\newblock Technical report, Technical Report. Cornel University, 2016.

\bibitem{ix}
Adam Belay, George Prekas, Ana Klimovic, Samuel Grossman, Christos Kozyrakis,
  and Edouard Bugnion.
\newblock $\{$IX$\}$: A protected dataplane operating system for high
  throughput and low latency.
\newblock In {\em 11th $\{$USENIX$\}$ Symposium on Operating Systems Design and
  Implementation ($\{$OSDI$\}$ 14)}, pages 49--65, 2014.

\bibitem{ar-grpc}
Rajarshi Biswas, Xiaoyi Lu, and Dhabaleswar~K Panda.
\newblock Accelerating tensorflow with adaptive rdma-based grpc.
\newblock In {\em 2018 IEEE 25th International Conference on High Performance
  Computing (HiPC)}, pages 2--11. IEEE, 2018.

\bibitem{scale-rpc}
Youmin Chen, Youyou Lu, and Jiwu Shu.
\newblock Scalable rdma rpc on reliable connection with efficient resource
  sharing.
\newblock In {\em Proceedings of the Fourteenth EuroSys Conference 2019},
  EuroSys ’19, New York, NY, USA, 2019. Association for Computing Machinery.

\bibitem{ni-arch}
Alexandros Daglis, Stanko Novakovi{\'c}, Edouard Bugnion, Babak Falsafi, and
  Boris Grot.
\newblock Manycore network interfaces for in-memory rack-scale computing.
\newblock {\em ACM SIGARCH Computer Architecture News}, 43(3S):567--579, 2015.

\bibitem{rpcvalet}
Alexandros Daglis, Mark Sutherland, and Babak Falsafi.
\newblock Rpcvalet: Ni-driven tail-aware balancing of $\mu$s-scale rpcs.
\newblock In {\em Proceedings of the Twenty-Fourth International Conference on
  Architectural Support for Programming Languages and Operating Systems}, pages
  35--48, 2019.

\bibitem{canns-consensus}
Huynh~Tu Dang, Pietro Bressana, Han Wang, Ki~Suh Lee, Hakim Weatherspoon, Marco
  Canini, Fernando Pedone, and Robert Soul{\'e}.
\newblock Network hardware-accelerated consensus.
\newblock {\em arXiv preprint arXiv:1605.05619}, 2016.

\bibitem{use-reads-or-not}
Aleksandar Dragojevic, Dushyanth Narayanan, and Miguel Castro.
\newblock Rdma reads: To use or not to use?
\newblock {\em IEEE Data Eng. Bull.}, 40(1):3--14, 2017.

\bibitem{farm}
Aleksandar Dragojevi{\'c}, Dushyanth Narayanan, Miguel Castro, and Orion
  Hodson.
\newblock Farm: Fast remote memory.
\newblock In {\em 11th $\{$USENIX$\}$ Symposium on Networked Systems Design and
  Implementation ($\{$NSDI$\}$ 14)}, pages 401--414, 2014.

\bibitem{via}
D.~{Dunning}, G.~{Regnier}, G.~{McAlpine}, D.~{Cameron}, B.~{Shubert},
  F.~{Berry}, A.~M. {Merritt}, E.~{Gronke}, and C.~{Dodd}.
\newblock The virtual interface architecture.
\newblock {\em IEEE Micro}, 18(2):66--76, 1998.

\bibitem{megapipe}
Sangjin Han, Scott Marshall, Byung-Gon Chun, and Sylvia Ratnasamy.
\newblock Megapipe: a new programming interface for scalable network i/o.
\newblock In {\em Presented as part of the 10th $\{$USENIX$\}$ Symposium on
  Operating Systems Design and Implementation ($\{$OSDI$\}$ 12)}, pages
  135--148, 2012.

\bibitem{xdp}
Toke H\o{}iland-J\o{}rgensen, Jesper~Dangaard Brouer, Daniel Borkmann, John
  Fastabend, Tom Herbert, David Ahern, and David Miller.
\newblock The express data path: Fast programmable packet processing in the
  operating system kernel.
\newblock In {\em Proceedings of the 14th International Conference on Emerging
  Networking EXperiments and Technologies}, CoNEXT ’18, page 54–66, New
  York, NY, USA, 2018. Association for Computing Machinery.

\bibitem{consensus-in-a-box}
Zsolt Istv{\'a}n, David Sidler, Gustavo Alonso, and Marko Vukolic.
\newblock Consensus in a box: Inexpensive coordination in hardware.
\newblock In {\em 13th $\{$USENIX$\}$ Symposium on Networked Systems Design and
  Implementation ($\{$NSDI$\}$ 16)}, pages 425--438, 2016.

\bibitem{mtcp}
EunYoung Jeong, Shinae Wood, Muhammad Jamshed, Haewon Jeong, Sunghwan Ihm,
  Dongsu Han, and KyoungSoo Park.
\newblock mtcp: a highly scalable user-level $\{$TCP$\}$ stack for multicore
  systems.
\newblock In {\em 11th $\{$USENIX$\}$ Symposium on Networked Systems Design and
  Implementation ($\{$NSDI$\}$ 14)}, pages 489--502, 2014.

\bibitem{netchain-switches}
Xin Jin, Xiaozhou Li, Haoyu Zhang, Nate Foster, Jeongkeun Lee, Robert
  Soul{\'e}, Changhoon Kim, and Ion Stoica.
\newblock Netchain: Scale-free sub-rtt coordination.
\newblock In {\em 15th $\{$USENIX$\}$ Symposium on Networked Systems Design and
  Implementation ($\{$NSDI$\}$ 18)}, pages 35--49, 2018.

\bibitem{e-rpc}
Anuj Kalia, Michael Kaminsky, and David Andersen.
\newblock Datacenter rpcs can be general and fast.
\newblock In {\em 16th {USENIX} Symposium on Networked Systems Design and
  Implementation ({NSDI} 19)}, pages 1--16, Boston, MA, February 2019. {USENIX}
  Association.

\bibitem{herd-kv}
Anuj Kalia, Michael Kaminsky, and David~G Andersen.
\newblock Using rdma efficiently for key-value services.
\newblock In {\em Proceedings of the 2014 ACM conference on SIGCOMM}, pages
  295--306, 2014.

\bibitem{rdma-design}
Anuj Kalia, Michael Kaminsky, and David~G Andersen.
\newblock Design guidelines for high performance $\{$RDMA$\}$ systems.
\newblock In {\em 2016 $\{$USENIX$\}$ Annual Technical Conference
  ($\{$USENIX$\}$$\{$ATC$\}$ 16)}, pages 437--450, 2016.

\bibitem{fasst-tr}
Anuj Kalia, Michael Kaminsky, and David~G Andersen.
\newblock Fasst: Fast, scalable and simple distributed transactions with
  two-sided ($\{$RDMA$\}$) datagram rpcs.
\newblock In {\em 12th $\{$USENIX$\}$ Symposium on Operating Systems Design and
  Implementation ($\{$OSDI$\}$ 16)}, pages 185--201, 2016.

\bibitem{tas-tcp}
Antoine Kaufmann, Tim Stamler, Simon Peter, Naveen~Kr. Sharma, Arvind
  Krishnamurthy, and Thomas Anderson.
\newblock Tas: Tcp acceleration as an os service.
\newblock In {\em Proceedings of the Fourteenth EuroSys Conference 2019},
  EuroSys ’19, New York, NY, USA, 2019. Association for Computing Machinery.

\bibitem{sift-cons}
Mikhail Kazhamiaka, Babar Memon, Chathura Kankanamge, Siddhartha Sahu, Sajjad
  Rizvi, Bernard Wong, and Khuzaima Daudjee.
\newblock Sift: resource-efficient consensus with rdma.
\newblock In {\em Proceedings of the 15th International Conference on Emerging
  Networking Experiments And Technologies}, pages 260--271, 2019.

\bibitem{r2p2-rpc}
Marios Kogias, George Prekas, Adrien Ghosn, Jonas Fietz, and Edouard Bugnion.
\newblock R2p2: Making rpcs first-class datacenter citizens.
\newblock In {\em 2019 $\{$USENIX$\}$ Annual Technical Conference
  ($\{$USENIX$\}$$\{$ATC$\}$ 19)}, pages 863--880, 2019.

\bibitem{ipipe-smart-nics}
Ming Liu, Tianyi Cui, Henry Schuh, Arvind Krishnamurthy, Simon Peter, and Karan
  Gupta.
\newblock Offloading distributed applications onto smartnics using ipipe.
\newblock In {\em Proceedings of the ACM Special Interest Group on Data
  Communication}, pages 318--333. 2019.

\bibitem{hadoop-rpc}
Xiaoyi Lu, Nusrat~S Islam, Md~Wasi-Ur-Rahman, Jithin Jose, Hari Subramoni, Hao
  Wang, and Dhabaleswar~K Panda.
\newblock High-performance design of hadoop rpc with rdma over infiniband.
\newblock In {\em 2013 42nd International Conference on Parallel Processing},
  pages 641--650. IEEE, 2013.

\bibitem{octopus}
Youyou Lu, Jiwu Shu, Youmin Chen, and Tao Li.
\newblock Octopus: an rdma-enabled distributed persistent memory file system.
\newblock In {\em 2017 $\{$USENIX$\}$ Annual Technical Conference
  ($\{$USENIX$\}$$\{$ATC$\}$ 17)}, pages 773--785, 2017.

\bibitem{snap}
Michael Marty, Marc de~Kruijf, Jacob Adriaens, Christopher Alfeld, Sean Bauer,
  Carlo Contavalli, Michael Dalton, Nandita Dukkipati, William~C Evans, Steve
  Gribble, et~al.
\newblock Snap: a microkernel approach to host networking.
\newblock In {\em Proceedings of the 27th ACM Symposium on Operating Systems
  Principles}, pages 399--413, 2019.

\bibitem{mellanox-guide}
Mellanox.
\newblock Rdma aware networks programming user manual.
\newblock
  \url{https://www.mellanox.com/related-docs/prod_software/RDMA_Aware_Programming_user_manual.pdf}.

\bibitem{pilaf-kv}
Christopher Mitchell, Yifeng Geng, and Jinyang Li.
\newblock Using one-sided $\{$RDMA$\}$ reads to build a fast, cpu-efficient
  key-value store.
\newblock In {\em Presented as part of the 2013 $\{$USENIX$\}$ Annual Technical
  Conference ($\{$USENIX$\}$$\{$ATC$\}$ 13)}, pages 103--114, 2013.

\bibitem{cell-kv}
Christopher Mitchell, Kate Montgomery, Lamont Nelson, Siddhartha Sen, and
  Jinyang Li.
\newblock Balancing $\{$CPU$\}$ and network in the cell distributed b-tree
  store.
\newblock In {\em 2016 $\{$USENIX$\}$ Annual Technical Conference
  ($\{$USENIX$\}$$\{$ATC$\}$ 16)}, pages 451--464, 2016.

\bibitem{so-numa}
Stanko Novakovic, Alexandros Daglis, Edouard Bugnion, Babak Falsafi, and Boris
  Grot.
\newblock Scale-out numa.
\newblock {\em ACM SIGPLAN Notices}, 49(4):3--18, 2014.

\bibitem{storm-tran}
Stanko Novakovic, Yizhou Shan, Aasheesh Kolli, Michael Cui, Yiying Zhang,
  Haggai Eran, Boris Pismenny, Liran Liss, Michael Wei, Dan Tsafrir, et~al.
\newblock Storm: a fast transactional dataplane for remote data structures.
\newblock In {\em Proceedings of the 12th ACM International Conference on
  Systems and Storage}, pages 97--108, 2019.

\bibitem{ram-cloud}
John Ousterhout, Arjun Gopalan, Ashish Gupta, Ankita Kejriwal, Collin Lee,
  Behnam Montazeri, Diego Ongaro, Seo~Jin Park, Henry Qin, Mendel Rosenblum,
  Stephen Rumble, Ryan Stutsman, and Stephen Yang.
\newblock The ramcloud storage system.
\newblock {\em ACM Trans. Comput. Syst.}, 33(3), August 2015.

\bibitem{affinity-accept}
Aleksey Pesterev, Jacob Strauss, Nickolai Zeldovich, and Robert~T Morris.
\newblock Improving network connection locality on multicore systems.
\newblock In {\em Proceedings of the 7th ACM european conference on Computer
  Systems}, pages 337--350, 2012.

\bibitem{arrakis}
Simon Peter, Jialin Li, Irene Zhang, Dan~RK Ports, Doug Woos, Arvind
  Krishnamurthy, Thomas Anderson, and Timothy Roscoe.
\newblock Arrakis: The operating system is the control plane.
\newblock {\em ACM Transactions on Computer Systems (TOCS)}, 33(4):1--30, 2015.

\bibitem{dare-cons}
Marius Poke and Torsten Hoefler.
\newblock Dare: High-performance state machine replication on rdma networks.
\newblock In {\em Proceedings of the 24th International Symposium on
  High-Performance Parallel and Distributed Computing}, pages 107--118, 2015.

\bibitem{zygos}
George Prekas, Marios Kogias, and Edouard Bugnion.
\newblock Zygos: Achieving low tail latency for microsecond-scale networked
  tasks.
\newblock In {\em Proceedings of the 26th Symposium on Operating Systems
  Principles}, pages 325--341, 2017.

\bibitem{dpdk}
The Linux~Foundation Projects.
\newblock Dataplane development kit.
\newblock \url{https://www.dpdk.org/}.

\bibitem{netmap}
Luigi Rizzo.
\newblock Netmap: a novel framework for fast packet i/o.
\newblock In {\em 21st USENIX Security Symposium (USENIX Security 12)}, pages
  101--112, 2012.

\bibitem{mercury}
Jerome Soumagne, Dries Kimpe, Judicael Zounmevo, Mohamad Chaarawi, Quincey
  Koziol, Ahmad Afsahi, and Robert Ross.
\newblock Mercury: Enabling remote procedure call for high-performance
  computing.
\newblock In {\em 2013 IEEE International Conference on Cluster Computing
  (CLUSTER)}, pages 1--8. IEEE, 2013.

\bibitem{memcached-rdma-kv}
Patrick Stuedi, Animesh Trivedi, and Bernard Metzler.
\newblock Wimpy nodes with 10gbe: leveraging one-sided operations in soft-rdma
  to boost memcached.
\newblock In {\em Presented as part of the 2012 $\{$USENIX$\}$ Annual Technical
  Conference ($\{$USENIX$\}$$\{$ATC$\}$ 12)}, pages 347--353, 2012.

\bibitem{da-rpc}
Patrick Stuedi, Animesh Trivedi, Bernard Metzler, and Jonas Pfefferle.
\newblock Darpc: Data center rpc.
\newblock In {\em Proceedings of the ACM Symposium on Cloud Computing}, SOCC
  ’14, page 1–13, New York, NY, USA, 2014. Association for Computing
  Machinery.

\bibitem{rfp}
Maomeng Su, Mingxing Zhang, Kang Chen, Zhenyu Guo, and Yongwei Wu.
\newblock Rfp: When rpc is faster than server-bypass with rdma.
\newblock In {\em Proceedings of the Twelfth European Conference on Computer
  Systems}, pages 1--15, 2017.

\bibitem{lite}
Shin-Yeh Tsai and Yiying Zhang.
\newblock Lite kernel rdma support for datacenter applications.
\newblock In {\em Proceedings of the 26th Symposium on Operating Systems
  Principles}, pages 306--324, 2017.

\bibitem{hydradb-kv}
Yandong Wang, Li~Zhang, Jian Tan, Min Li, Yuqing Gao, Xavier Guerin, Xiaoqiao
  Meng, and Shicong Meng.
\newblock Hydradb: a resilient rdma-driven key-value middleware for in-memory
  cluster computing.
\newblock In {\em SC'15: Proceedings of the International Conference for High
  Performance Computing, Networking, Storage and Analysis}, pages 1--11. IEEE,
  2015.

\bibitem{drtm+h}
Xingda Wei, Zhiyuan Dong, Rong Chen, and Haibo Chen.
\newblock Deconstructing rdma-enabled distributed transactions: Hybrid is
  better!
\newblock In {\em 13th $\{$USENIX$\}$ Symposium on Operating Systems Design and
  Implementation ($\{$OSDI$\}$ 18)}, pages 233--251, 2018.

\bibitem{drtm-tr}
Xingda Wei, Jiaxin Shi, Yanzhe Chen, Rong Chen, and Haibo Chen.
\newblock Fast in-memory transaction processing using rdma and htm.
\newblock In {\em Proceedings of the 25th Symposium on Operating Systems
  Principles}, pages 87--104, 2015.

\bibitem{gram}
Ming Wu, Fan Yang, Jilong Xue, Wencong Xiao, Youshan Miao, Lan Wei, Haoxiang
  Lin, Yafei Dai, and Lidong Zhou.
\newblock Gram: Scaling graph computation to the trillions.
\newblock In {\em Proceedings of the Sixth ACM Symposium on Cloud Computing},
  pages 408--421, 2015.

\bibitem{dist-learn}
Jilong Xue, Youshan Miao, Cheng Chen, Ming Wu, Lintao Zhang, and Lidong Zhou.
\newblock Fast distributed deep learning over rdma.
\newblock In {\em Proceedings of the Fourteenth EuroSys Conference 2019}, pages
  1--14, 2019.

\bibitem{dist-tree}
Tobias Ziegler, Sumukha Tumkur~Vani, Carsten Binnig, Rodrigo Fonseca, and Tim
  Kraska.
\newblock Designing distributed tree-based index structures for fast
  rdma-capable networks.
\newblock In {\em Proceedings of the 2019 International Conference on
  Management of Data}, pages 741--758, 2019.

\end{thebibliography}

\end{document}